\documentclass[%
 reprint,
superscriptaddress,
 amsmath,amssymb,
 aps,
]{revtex4-2}

\usepackage{graphicx}
\usepackage{dcolumn}
\usepackage{bm}
\usepackage[caption=false]{subfig}
\usepackage[usenames,dvipsnames,svgnames,table]{xcolor}

\usepackage[final]{changes}
\definechangesauthor[name={Francesco Papoff}, color=blue]{fp}
\definechangesauthor[name={Mark Anthony Carroll}, color=red]{mc}
\definechangesauthor[name={Giampaolo D'Alessandro}, color=OliveGreen]{gd}
\definechangesauthor[name={Gian Luca Lippi}, color=violet]{gll}
\definechangesauthor[name={Gian-Luca Oppo}, color=orange]{glo}

\begin{document}

\preprint{APS/123-QED}

\title{Thermal, quantum anti-bunching and lasing thresholds from single  emitters to macroscopic devices}
\author{Mark Anthony Carroll}
\affiliation{Department of Physics, University
of Strathclyde,  107 Rottenrow,  Glasgow G4 0NG, UK.}
\author{Giampaolo D'Alessandro}
\affiliation{School of Mathematics, University of Southampton, Southampton SO17 1BJ, United Kingdom}
\author{Gian Luca Lippi}
\affiliation{Universit\'e C\^ote d'Azur, Institut de Physique de Nice, UMR 7710 CNRS, 1361 Route des Lucioles, 06560 Valbonne, France}
\author{Gian-Luca Oppo}\affiliation{Department of Physics, University
of Strathclyde,  107 Rottenrow,  Glasgow G4 0NG, UK.}
\author{Francesco Papoff}
\email{f.papoff@strath.ac.uk}
\affiliation{Department of Physics, University
of Strathclyde,  107 Rottenrow,  Glasgow G4 0NG, UK.}

\date{\today}

\begin{abstract} 
Starting from a fully quantized Hamiltonian for an ensemble of identical
emitters coupled to the modes of an optical cavity, we determine analytically   regimes of thermal, collective anti-bunching  and laser emission that depend explicitly on the number of emitters. The lasing regime is reached for a number of emitters above a critical number -- which depends on the light-matter coupling, detuning and the dissipation rates -- via a universal transition from thermal emission to collective anti-bunching to lasing as the pump increases. Cases where the second order intensity correlation fails to predict laser action are also presented.
\end{abstract}

\maketitle

Optical cavities containing emitters with discrete energy levels such as atoms, ions or quantum dots, have proved extremely effective both as a tool to investigate fundamental properties of light-matter interaction and as a way to produce light with engineered statistical properties. Historically, the earliest optical cavities contained very large numbers of emitters to overcome losses and occupied macroscopic volumes. Significant  improvements in cavity quality have led in recent years to micro and nano cavities~\cite{Hill2014} that provide detectable fields even with very few emitters and
promise commercial applications with uses ranging from components on integrated circuits to medicine~\cite{Ma2019}. Their small size comes with the benefit of lower energy consumption and increased energy efficiency, making them attractive for extreme miniaturization. Although the basic quantum interaction process between the cavity modes and the  emitters is the same for all cavities,
different approximations of the expectation values of light-matter interaction and photon number (correspondiong to the classical intensity) lead to two  classes of models: one for micro and nano systems, the other for macroscopic systems. In the latter, e.g. the Maxwell-Bloch semi-classical models for macroscopic systems~\cite{Narducci1988}, the only expectation values considered are those of the emitters' \added[id=fp]{\added[id=gd]{raising} and lowering operators,  representing excitation and de-excitation of an electron, and of the cavity mode creation and destruction operators,  representing emission and absorption of a photon.  The  expectation values of the destruction and creation operators correspond to the complex amplitude of the classical coherent field and its complex conjugate. Similarly, the expectation values of \added[id=gd]{raising} and lowering operators correspond to the amplitude of the 
medium polarization and its complex conjugate.} Macroscopic models neglect correlations  among these operators,  both in the intensity and in the photon-matter coupling, also called the photon-assisted polarization. This approximation allows one to predict the threshold for laser emission, but is not suitable for the analysis of non-lasing emission \added[id=fp]{because photon-assisted polarization is essential to model correctly spontaneous emission}~\cite{kira2011semiconductor}. Quantum models for micro and nano lasers take the exact opposite approach: they consider only the correlations and neglect the expectation values~\cite{gies07a, chow14a, kreinberg17a, kira2011semiconductor} \added[id=fp]{corresponding to the amplitudes of the classical coherent field and polarization}. With this approximation non-lasing emission can be modelled, but it is not possible to identify the onset of lasing. The same problem  also affects rate equation models of micro- and nanolasers that add average spontaneous emission to coherent emission~\cite{Yokoyama1989, rice1994photon, mork2018rate}. A key parameter related to the size of the system is the spontaneous emission factor, $\beta$, defined as the ratio of the spontaneous emission rate into the lasing mode to the total spontaneous emission. $\beta^{-1}$, proportional to the number of electromagnetic modes in the cavity volume, thus characterises the system size.  In macroscopic systems $\beta \ll 1$, while $\beta=1$ is the nanoscale  limit, in which only the lasing mode remains accessible to spontaneous and stimulated emission. For this value of $\beta$ the output power linearly follows the input power, and for this reason this laser is considered ``thresholdless''~\cite{Yokoyama1989,Ning2013}. This regime poses questions on how to define the laser threshold~\cite{rice1994photon} and identify \added[id=gll]{coherent}  emission~\cite{samuel09a}, which we address unambiguously in this paper.

Starting from a fully quantized Jaynes-Cummings Hamiltonian in the Heisenberg picture, \added[id=fp]{ see Supplementary Material (SM) Eq.(1)}, we derive a model for the emission of any number of identical two-level emitters coupled to one mode of the cavity ~\cite{fricke96a,feldtmann06a}. However, we do not approximate the expectation values of intensity and light-matter interaction,  \added[id=fp]{including}  the variables of both macroscopic and nano lasers, and apply dynamical system methods \added[id=fp]{defined for systems of any dimension}~\cite{solari96} to identify the laser threshold. Analytical solutions seamlessly connect
single emitter devices with devices containing millions of
emitters, predicting where the thermal, quantum and coherent emission regimes lie with respect to one another. While the extent and existence of these regions in the space of parameters depend on the number of emitters, we find two universal features that are common to all lasing devices. The first is that lasing, when possible, is reached  via  a universal sequence of transitions as the pump is increased. The  emission of the non-lasing state evolves continuously from thermal (with second order intensity correlation $1< g^{(2)}(0)\le 2$) to anti-bunching (with $0 \le  g^{(2)}(0) < 1$) until the laser threshold is crossed and the non-lasing state becomes unstable. A coherent laser field, due to a lasing collective state, appears at this threshold and its amplitude increases as a function of the pump.  This is the same instability predicted for macroscopic laser by Maxwell-Bloch models. However, in these models the  total field before the instability is zero, while in our theory  before the threshold only the coherent field is zero, while the incoherent field is non zero. The second universal feature of lasers is that the emerging coherent field has a well defined frequency: our model shows that neither the number of emitters nor the effective cavity volume ($\beta$ factor) influence the frequency value, which remains a general feature of the cavity-emitter interaction.   We give examples \added[id=gll]{where the measurement}  of $g^{(2)}(0)$ cannot identify laser emission.

We consider \added[id=fp]{identical} emitters with two energy levels and
one electron, and assume that all transitions conserve the electron spin.  
This model applies 
to atoms and ions with suitable energy level structure, and also to
shallow quantum dots (possessing two localized levels) at temperatures low enough to neglect Coulomb and phonon interactions.
\added[id=fp]{Assuming that detuning and coupling coefficients  with the mode are identical~\cite{kreinberg17a,moody18a} is justified by numerical simulations for emitters with $10\%$ random variations of detuning and coupling coefficients.  In these simulations all emitters' variables --
\added[id=gd]{started from} random initial conditions -- after a transient converge to common values that match extremely well those obtained  using the same parameters and expectation values for all emitters, see SM Fig. 2. }Since current technology enables very accurate positioning \cite{Kaganskiy2019}, the range of fluctuations that we are considering is quite realistic
and covers estimates
  which can be expected from QD positions inside typical cavity
realizations and other experimental
fluctuations.

In the following $c^\dagger$, $v^\dagger$, $b^\dagger$ ($c$, $v$, $b$) are operators that create (annihilate), respectively, an electron in the upper  energy level (conduction level in quantum dot terminology),  
in the lower energy level (valence level), and a photon in the laser mode. 
We take the expectation values of the Heisenberg equations for   each operator in the Hamiltonian, truncating the resulting infinite hierarchy
of coupled differential equations by keeping only correlation functions that appear in the cluster expansion of the Hamiltonian~\cite{fricke96a, kira2011semiconductor}. 
The variables of the model are the expectation values of carrier population, $\langle c^\dagger c\rangle$, \added[id=fp]{ mode destruction operator}, $\langle b \rangle$\added[id=gll]{,} and \added[id=fp]{  emitter lowering operator}, $\langle v^\dagger c\rangle$, and the intensity and photon-polarization correlations, $\delta\langle b^\dagger b\rangle$ and $\delta\langle b c^\dagger v\rangle$.
These correlations appear through the cluster expansions  $\langle b c^\dagger v \rangle = \delta \langle b c^\dagger v \rangle + \langle b \rangle \langle c^\dagger v \rangle$ and
 $\langle b^\dagger b \rangle = \delta \langle b^\dagger b \rangle +  \langle b^\dagger \rangle \langle b \rangle$ of terms in the Hamiltonian. 
\added[id=fp]{  Note that  the equations for $\langle b^\dagger  \rangle$, $\langle c^\dagger v \rangle$, $\langle b^\dagger v^\dagger c \rangle$ are the complex conjugated of the equation for $\langle  b \rangle$, $\langle v^\dagger c \rangle$, $\langle b c^\dagger v \rangle$, so they are not explicitly included in the model.}
 The model equations are

\begin{align}
  d_t \langle c^\dagger c\rangle =
  &-(\gamma_{nl} + \gamma_{nr})\langle c^\dagger c \rangle + r(1 - \langle c^\dagger c\rangle) \nonumber \\
  &- 2\mathfrak{Re}[g(\delta\langle b c^\dagger v\rangle + \langle b \rangle\langle c^\dagger v\rangle)], \label{mod_eq1} \\
  d_t \delta\langle b c^\dagger v\rangle =
  &-(\gamma_c + \gamma +i\Delta \nu)\delta\langle b c^\dagger v \rangle + g^*[\langle c^\dagger c \rangle \nonumber \\
  &+ \delta\langle b^\dagger b\rangle(2\langle c^\dagger c\rangle - 1) - \vert\langle v^\dagger c\rangle\vert^2 ], \\
  d_t  \delta\langle b^\dagger b\rangle =
  &-2\gamma_c \delta\langle b^\dagger b\rangle + 2N\mathfrak{Re}(g \delta\langle b c^\dagger v\rangle),\label{mod_eq3} \\
  d_t  \langle b\rangle =
  &-(\gamma_c + i\nu)\langle b \rangle + g^*N\langle v^\dagger c\rangle, \label{mod_eq4} \\
  d_t  \langle v^\dagger c\rangle =
  &-(\gamma + i\nu_\varepsilon)\langle v^\dagger c\rangle + g[ \langle b \rangle(2\langle c^\dagger c \rangle - 1)], \label{mod_eq5}
\end{align}

where \added[id=fp]{ $r$} is the  pump \added[id=gll]{rate} per emitter (the \added[id=gll]{one for} the whole system is $N r$), \added[id=fp]{with non resonant pump photons absorption and carrier transport effects approximated via an injection rate~\cite{kreinberg17a}}. $\nu$ is the frequency of  the resonant mode and $\nu_\varepsilon$ is the resonant frequency of  emitters, $\Delta \nu =   \nu_\varepsilon - \nu$ is the detuning between the field and emitter, $g$ the light-matter coupling strength and $N$ the number of emitters. The decay rates for the laser mode, $\gamma_c$, the population, $\gamma_{nr}$, and polarisation, $\gamma$, introduce dissipation. The dissipative part of these equations can be obtained by considering Lindblad terms describing the coupling to a Markovian bath~\cite{florian13a, leymann14a}.  \added[id=fp]{$\gamma_{nr}$ is due to} \added[id=mc]{non-radiative transitions}, while the additional decay rate
for the population, $\gamma_{nl}$, results from the adiabatic elimination of rapidly decaying non lasing modes~\cite{chow13a}. The spontaneous emission factor, $\beta$, is related to the cavity losses by $\beta = \frac{\gamma_l}{\gamma_{nl} + \gamma_l}$, where $\gamma_l$ is the rate of spontaneous emission into the lasing mode. The expectation value of lower level population, $\langle v^\dagger v\rangle $ has been eliminated using $\langle c^\dagger c\rangle + \langle v^\dagger v\rangle = 1$. \added[id=gd]{The variables $\langle b \rangle$ and $\langle v^\dagger c \rangle$ correspond to the amplitudes of the coherent field and the medium polarization of semi-classical models of macroscopic lasers and were previously neglected in nanolasers. The imaginary coefficients in their the equations show that they have fast oscillations with frequency of the order of that of the cavity mode. Hence,} we call them ``fast'' and the remaining variables ``slow''.

\begin{figure}
  \includegraphics[width=0.49\columnwidth]{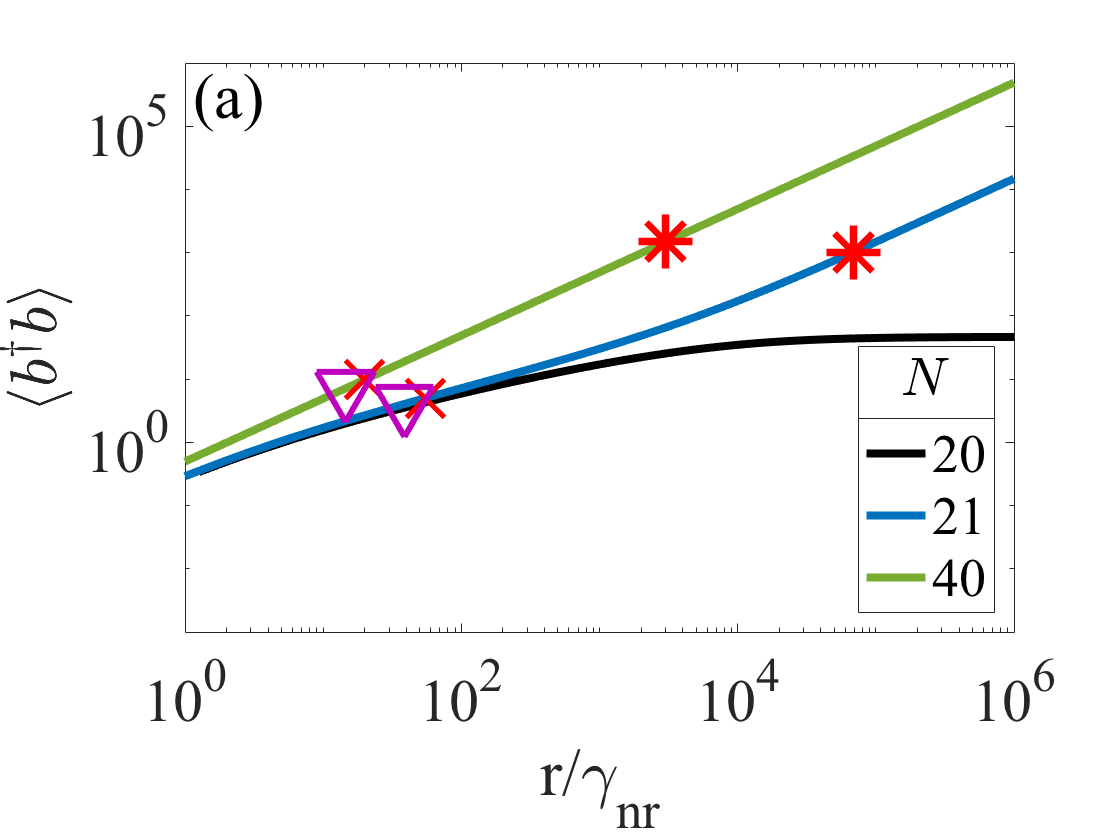}
  \includegraphics[width=0.49\columnwidth]{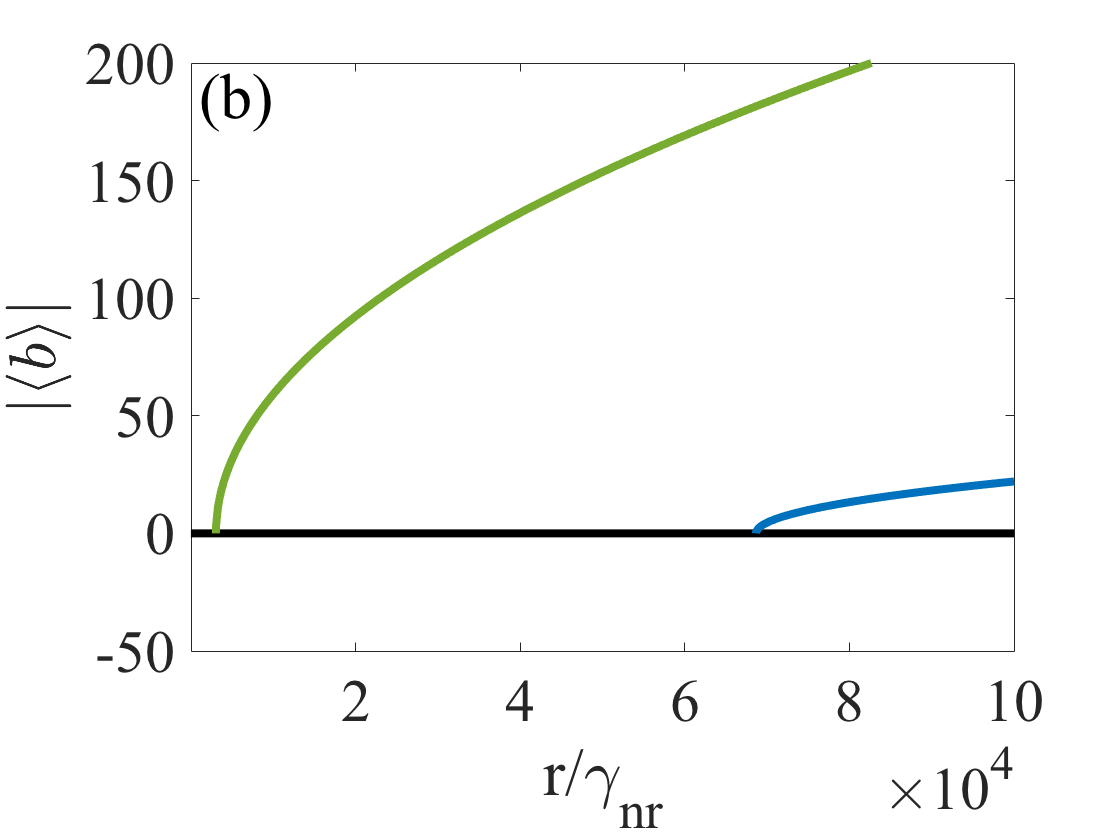}
  \includegraphics[width=0.49\columnwidth]{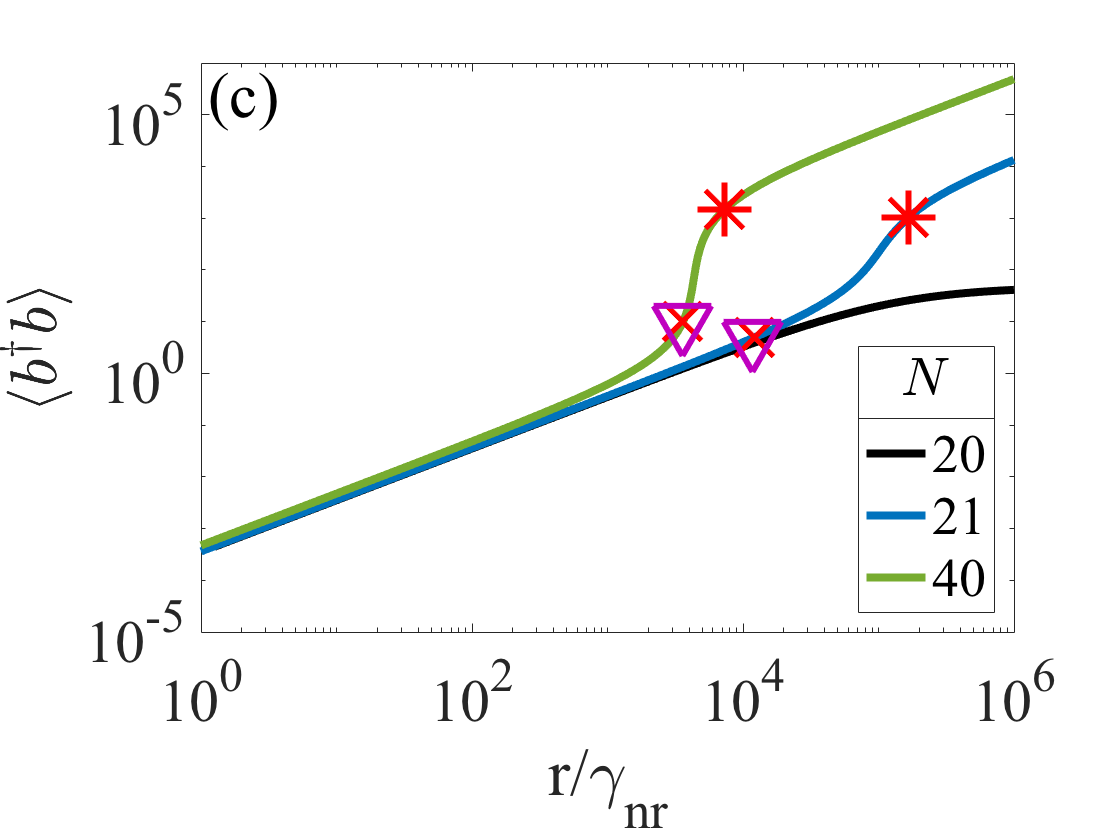}
  \includegraphics[width=0.49\columnwidth]{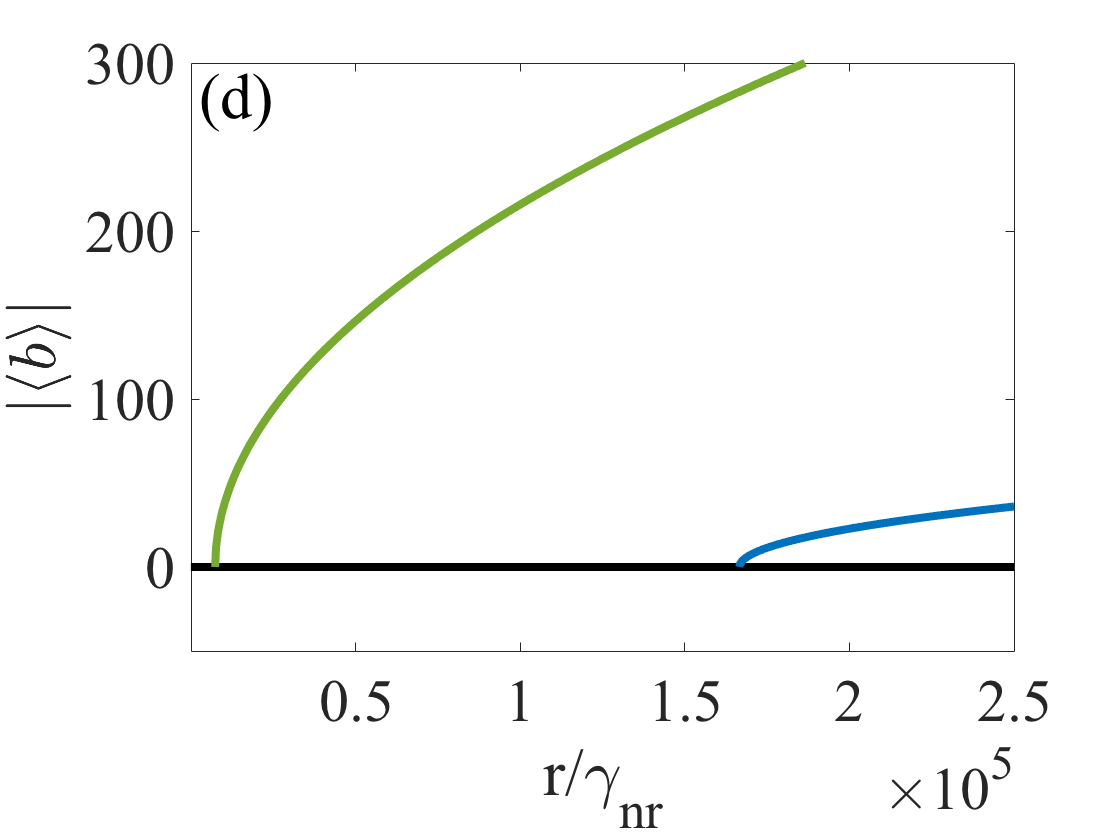}
\caption{ I/O curves: Intensity, \added[id=gd]{$\langle b^\dagger b \rangle = \delta \langle b^\dagger b \rangle +  \langle b^\dagger \rangle \langle b \rangle$,} and coherent field  amplitude for different $N$ above and below $N_c$ \added[id=mc]{versus} pump for $\beta=1$ ($\gamma_{nl}=0$), (a) and (b);  and $\beta=7\times10^{-4}$ ($\gamma_{nl}=1.4\times10^{12}s^{-1}$ and $\gamma_{l}=9.68\times10^{8}s^{-1}$), (c) and (d). \added[id=mc]{Laser and anti-bunching thresholds are marked by red stars and crosses, respectively.} The purple triangles mark the anti-bunching threshold where we include higher order correlations (this shows our approximation to be valid). Note that a laser threshold is found even for a ``thresholdless" device (green line \added[id=gll]{in (a)}). Parameters: $\gamma=10^{13}s^{-1}$, $\gamma_c=10^{10}s^{-1}$, $\gamma_{nr}=10^9s^{-1}$, and $g=7\times10^{10}s^{-1}$. Losses are kept constant in all figures.    }\label{fig:Figure1}

\end{figure}

The non-lasing stationary state is found by setting to zero the fast variables and the time derivatives in Eqs.(\ref{mod_eq1}-\ref{mod_eq3}). This state exists for all values of the control parameters and its emission evolves continuously from thermal to anti-bunching. The boundary between these two regimes is identified by the curve $g^{(2)}(0)=1$. We find a very good analytical approximation of the $g^{(2)}(0)=1$ curve assuming that the correlations used here are independent of higher order correlations~\cite{chow14a}, so that $g^{(2)}(0)\sim 2+A/B$, \added[id=fp]{see SM Eqs.(12-14)}. 
We determine the stability of the non-lasing state by deriving the equations that govern the evolution of small perturbations in the linear regime. We find that the perturbations of the fast variables are decoupled from the perturbations of the slow variables allowing us to  determine analytically the stability of the non-lasing state (see SM Eqs.(22-24)). The non-lasing state is stable for $\langle c^\dagger c\rangle <\langle c^\dagger c\rangle_{th}$ and unstable for  $\langle c^\dagger c\rangle >\langle c^\dagger c\rangle_{th}$ where
\begin{equation}
    \label{Cc_th}
  \langle c^\dagger c\rangle_{th} = \frac{1}{2}
  + \frac{\gamma_c\gamma}{2 N\vert g \vert^{2}}
  \left [ 1 +
    \left ( \frac{\Delta \nu}{\gamma_c + \gamma} \right )^{2}
  \right ]
\end{equation}
is the laser threshold. 
Because $\langle c^\dagger c\rangle<1$, lasing can only happen when the number of emitters satisfies the condition
\begin{equation}
    \label{N}
  \frac{\gamma_c\gamma}{\vert g\vert^2} \left [
    1+ \left ( \frac{\Delta \nu }{\gamma_c + \gamma} \right )^{2}
  \right ] \leq N. 
\end{equation}
\added[id=fp]{ The laser frequency is determined using trial solutions $\langle b \rangle, \langle v^\dagger c \rangle \propto e^{-i\Omega t}$
in Eqs.(\ref{mod_eq4},\ref{mod_eq5}) and is}
\begin{equation}
  \label{eq_omega}
  \Omega = \nu + \frac{\gamma_c \Delta \nu}{\gamma_c + \gamma}. 
\end{equation}
A few points are worth highlighting. First, the threshold can be calculated for any values of the decay rate $\gamma_{nl}$, including the so called ``thresholdless" case $\gamma_{nl}=0$. Second, neither $\langle c^\dagger c\rangle_{th}$ nor the critical number of emitters necessary to lase depend on $\beta$. However, the value of the pump\added[id=gll]{ per emitter} required to reach $\langle c^\dagger c\rangle_{th}$ depends on $\beta$ and decreases as $\beta$ increases, see SM Eq.(20) for the analytic expression for $r$ at laser threshold. Third, the laser frequency $\Omega$ is independent of the number of emitters, $N$. 

\begin{figure}
  \includegraphics[width=0.49\columnwidth]{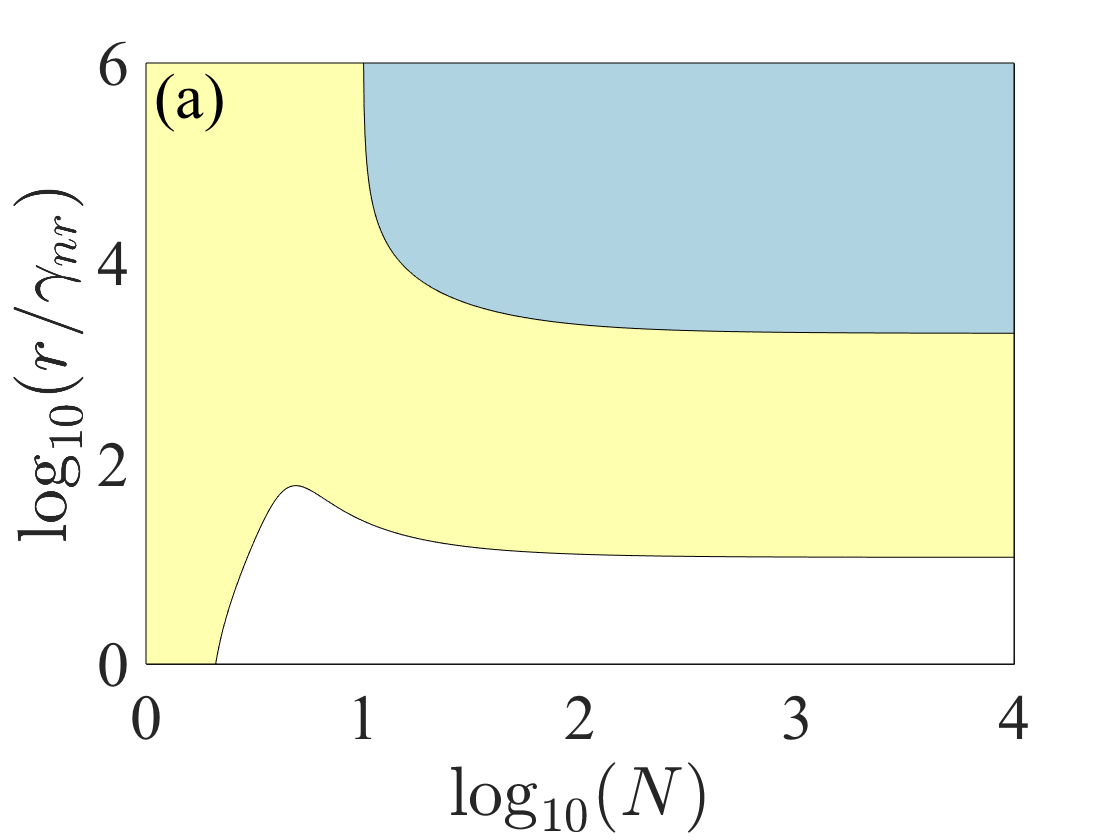}
  \includegraphics[width=0.49\columnwidth]{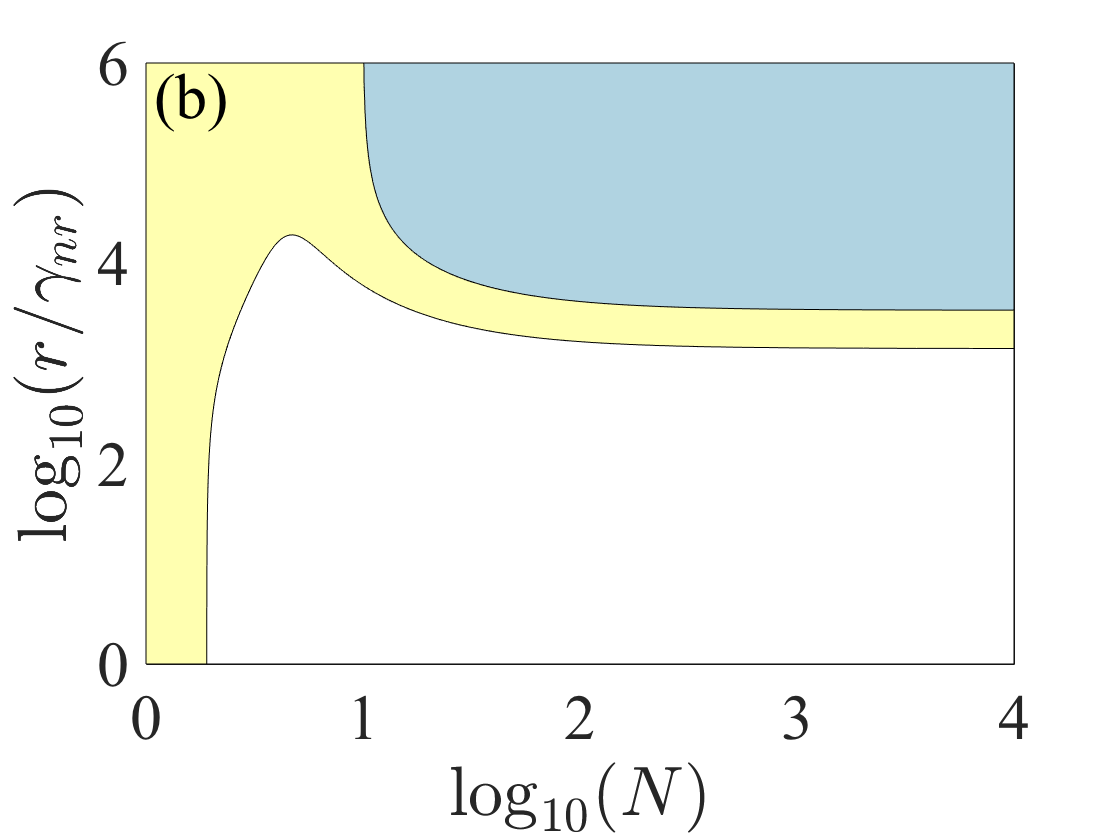}
  \includegraphics[width=0.49\columnwidth]{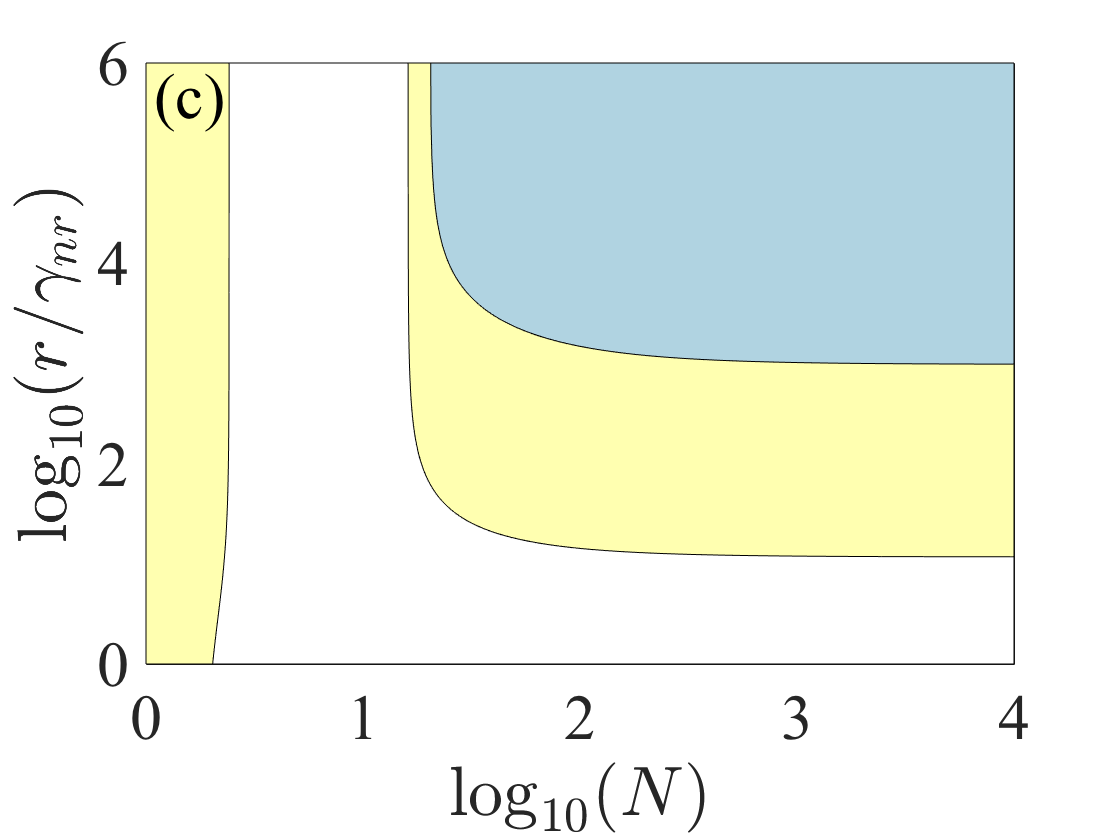}
  \includegraphics[width=0.49\columnwidth]{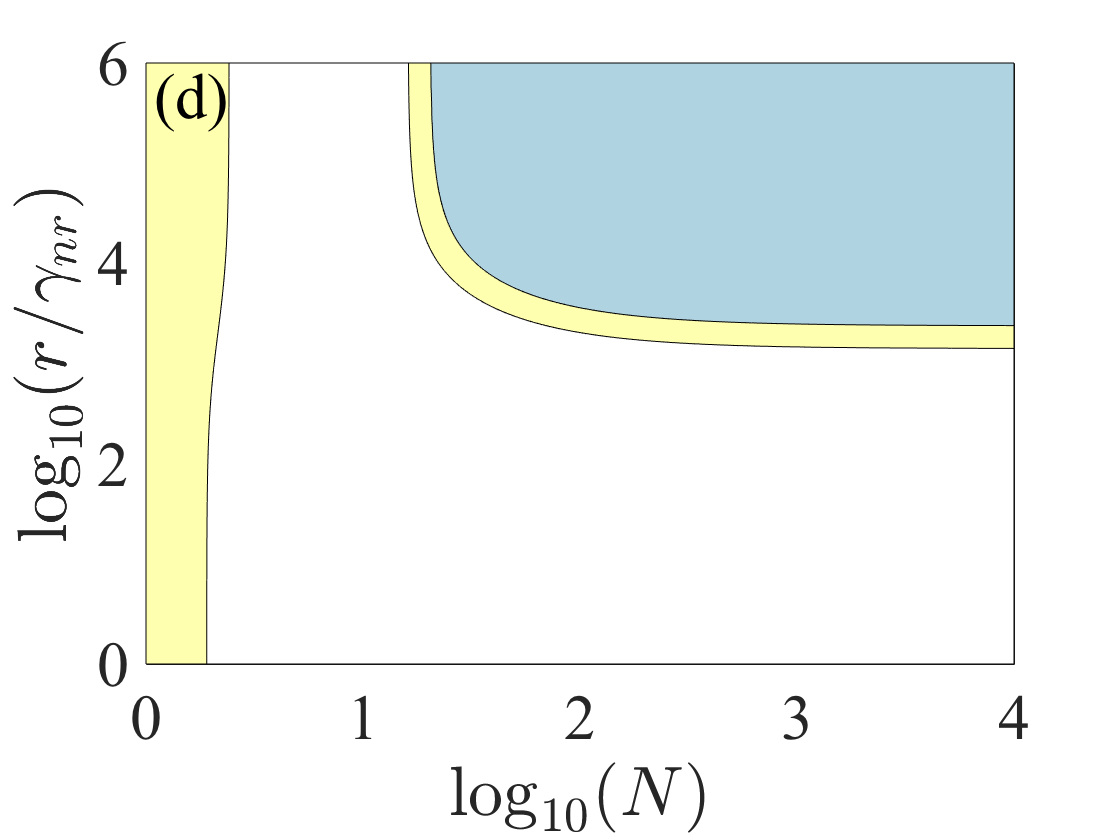}
 \caption{Phase Diagrams: thermal, anti-bunching and lasing regimes correspond to the white, yellow and blue regions, respectively. Where $\beta=1$ and $g=10^{11}s^{-1}$ in (a); $\beta=7\times10^{-4}$ and $g=10^{11}s^{-1}$ in (b); $\beta=1$ and $g=7\times10^{10}s^{-1}$ in (c); $\beta=7\times10^{-4}$ and $g=7\times10^{10}s^{-1}$ in (d). For large $g$ the thermal region that extends to large pump values vanishes, and for a lasing a device with $\beta=1$ the anti-bunching regime exists over greater values of the pump. }\label{fig:antibunching-laser}

\end{figure}

In Fig.~\ref{fig:Figure1}a and Fig.~\ref{fig:Figure1}b the intensity and coherent field versus pump for a $\beta=1$ device are shown. The different number of emitters correspond to values above (blue and olive curves) and below (black curves) the critical number, $N_c$, required for a laser using Eq.(7).  Below $N_c$, the minimum integer that satisfies Eq.\eqref{N}, the coherent field amplitude is always zero and  the intensity saturates at high pump. Above $N_c$\added[id=gll]{,} $\langle b \rangle \ne 0$ emerges through a pitchfork bifurcation.  The emergence of a pitchfork bifurcation also coincides with the growth of the intensity and a clear qualitative difference is apparent between the curves corresponding to lasing and non lasing devices.  The I/O curve for 40 quantum dots (Fig.~\ref{fig:Figure1}a) illustrates the impossible task of determining  the laser threshold  for a $\beta=1$ device, identifiable only through the fast variables. \added[id=fp]{ Comparison of these thresholds with standard estimates is given in  SM Fig. 4.} Notice that all graphs plot pump for a single quantum dot.  Thus, comparison between devices with different $N$ requires multiplication of each horizontal scale by $N$. Variations in the I/O similar to those shown in Fig.~\ref{fig:Figure1}a for $N=21$ and $N=40$ can be obtained with the same value of N and changing the detuning. This is observed in experiments where detuning decreases the effective number of quantum dots interacting with the field. \added[id=mc]{See SM Fig. 6 for the effect of detuning}. Experiments have obtained this kind of I/O response through cavity or thermal tuning~\cite{Takiguchi2016,Ota2017,Jagsch2018}.  While so far explained only through ad-hoc calculations, here the continuous transformation with change in characteristic I/O shape emerges thanks to self-consistent modeling.

Fig.~\ref{fig:Figure1}c and Fig.~\ref{fig:Figure1}d  show the intensity and coherent field for a device where $\beta \ll 1$. One clear difference is that \added[id=gll]{the} intensity profile is no longer linear and there is the characteristic s-shaped I/O curve. Notable is the appearance of the bifurcation at the knee of the upper branch of the I/O curve, rather than at the inflection point (as from~\cite{rice1994photon}). This points to a substantial contribution from the incoherent emission to the intensity growth before the bifurcation and highlights the difficulty intrinsic in determining threshold through the I/O curve. 

\begin{figure}
  \includegraphics[width=0.49\columnwidth]{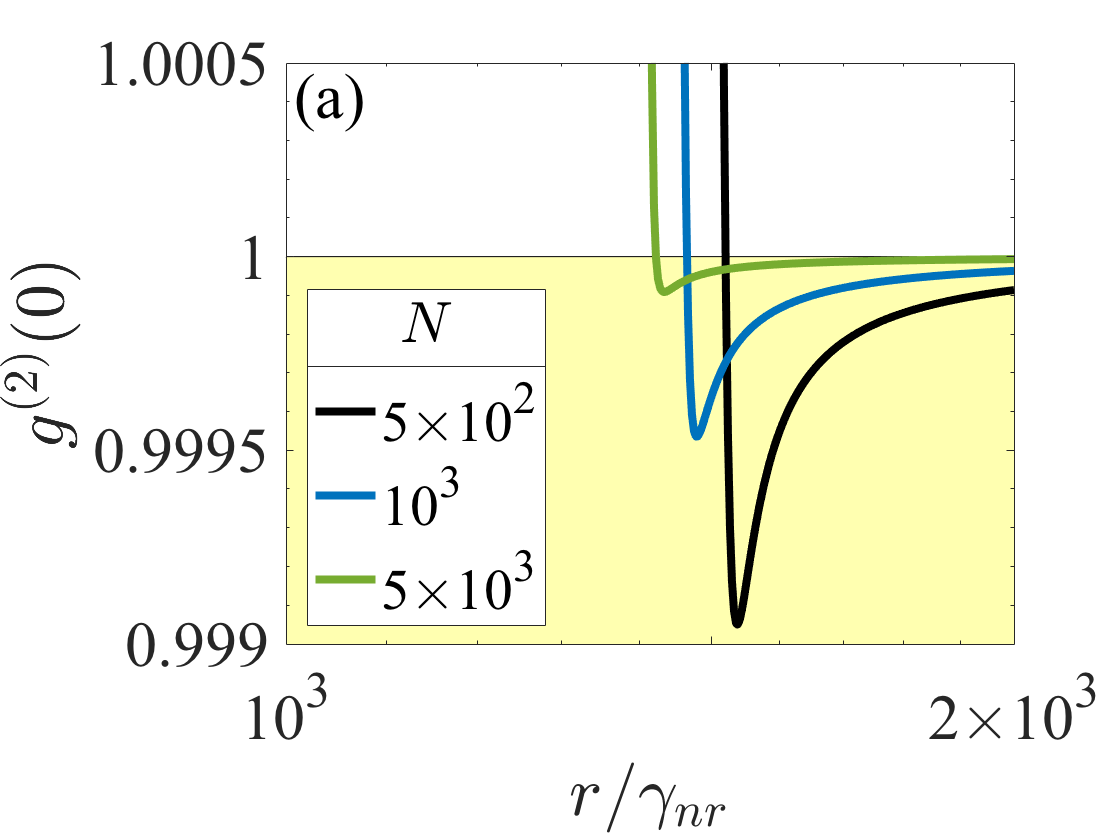}
  \includegraphics[width=0.49\columnwidth]{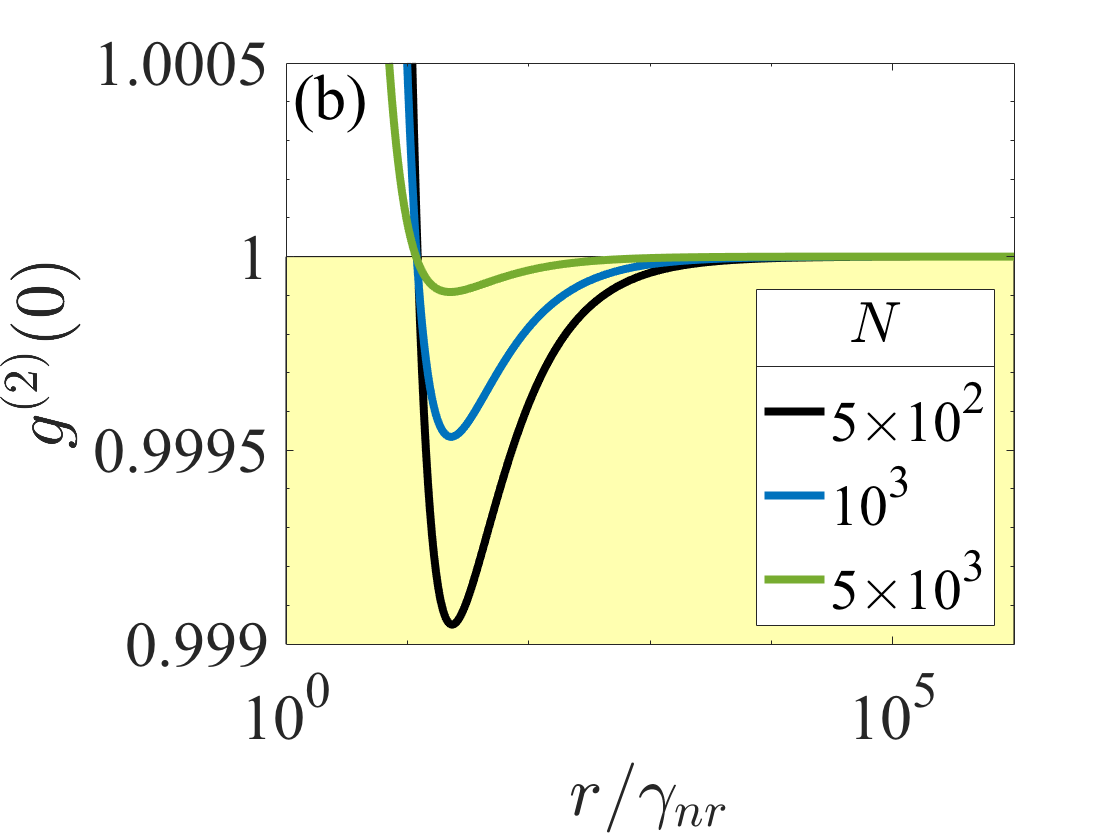}
  \includegraphics[width=0.49\columnwidth]{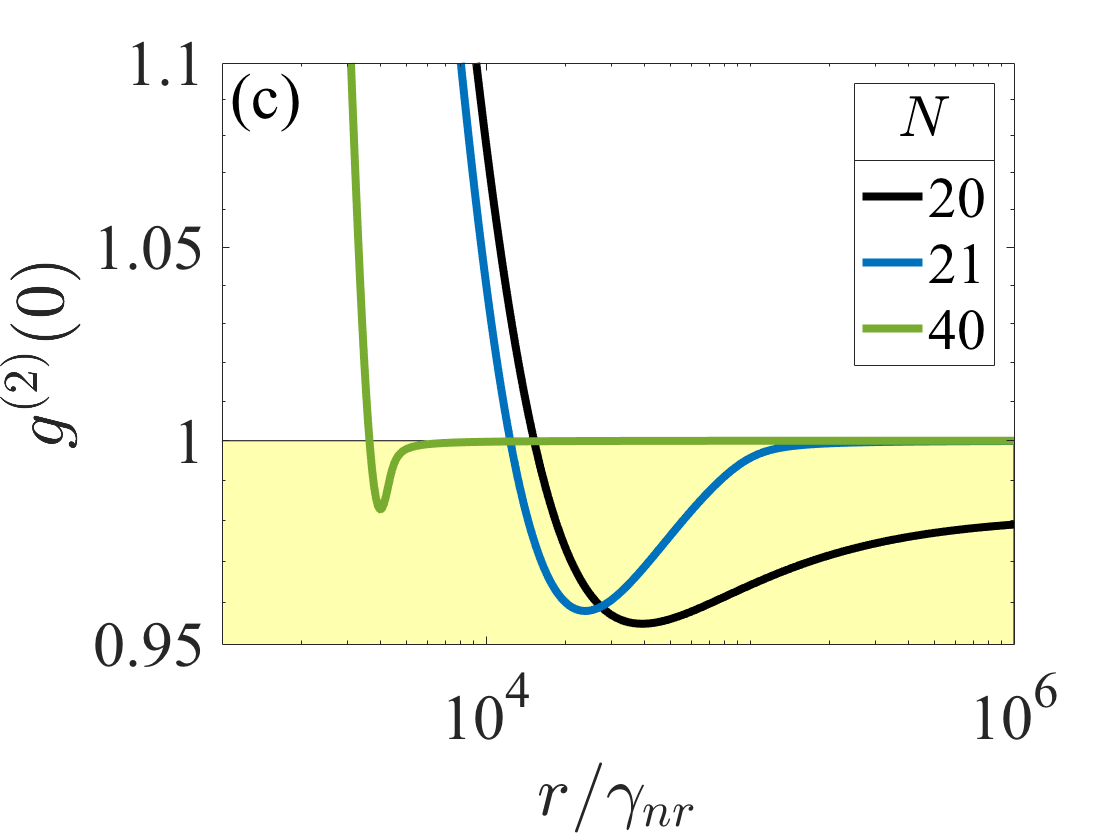}
  \includegraphics[width=0.49\columnwidth]{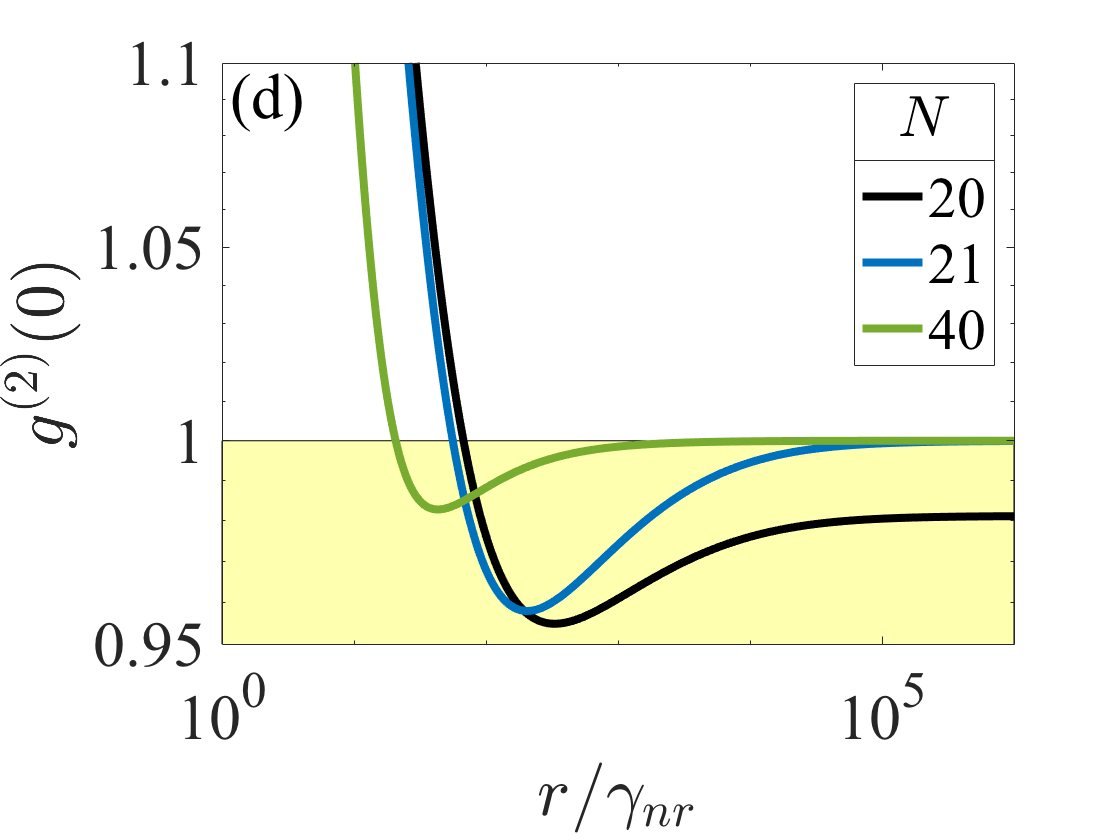}
\caption{$g^{(2)}(0)$ versus pump for different $N$ and $\beta$ where the coloured regions correspond to the same in Fig.~\ref{fig:antibunching-laser}. In (a) and (c) $\beta=7\times10^{-4}$; and in (b) and (d) $\beta=1$. Note the large range of pump values in (a) and (b) where $g^{(2)}(0)$ is smaller, but very close to $1$, making $g^{(2)}(0)$ difficult to use as an experimental indicator of lasing. All curves in (a) converge to the same value for larger pump values not shown. All curves are obtained for $g=7\times10^{10}s^{-1}$ thus lasing occurs for $N=21$. }\label{fig:g2-vs-pump}

\end{figure}

Fig.~\ref{fig:antibunching-laser} shows the analytic solutions of $g^{(2)}(0)=1$ and the laser threshold as functions of the pump and $N$ for different values of $g$ and $\beta$. 
For a device capable of lasing there is a well defined path of emission as the pump increases (vertical cut in the graph): from thermal to anti-bunching; then from anti-bunching to lasing. This is independent of the value of $\beta$. Fig.~\ref{fig:antibunching-laser}a and Fig.~\ref{fig:antibunching-laser}c show an anti-bunching regime that exists for a large region of the pump and extends from nano- to macroscopic lasers with $\beta=1$.  Anti-bunching has been observed in an experiment with a high cavity Q micropillar containing $N \approx 15$ pumped quantum dots~\cite{Wiersig2009} and in numerical simulations of quantum dot nanocavities~\cite{chow14a},\cite{gies15a}. The observation is consistent with our prediction (Fig.~\ref{fig:g2-vs-pump}), which shows stronger anti-bunching close to $N_c$; an increase in $N$ reduces the range and amplitude, thus making an experimental observation much more difficult (no anti-bunching was in fact observed upon doubling of the quantum dot number~\cite{Wiersig2009}).    Fig.~\ref{fig:g2-vs-pump}a shows
the three regimes for lasers with $\beta=7\times10^{-4}$. Although anti-bunching is present in macroscopic lasers, $g^{(2)}(0)$ is so close to unity that it becomes very difficult to distinguish between a laser and a device emitting anti-bunched light, thus stressing that $g^{(2)}(0)$ is not a sufficient indicator of laser action. \added[id=mc]{We show in Fig. 3 of the SM a comparison of Eqs. (1-5) with the master equation given in \cite{rice1994photon} and a cQED model  \cite{kreinberg17a}.} 

In conclusion, we have derived a model that includes coherent and incoherent emission and derived analytically thresholds that separate thermal, anti-bunching and lasing emission regimes from single emitter to macroscopic devices.  We predict the qualitative differences between the I/O curves of the photon number above and below threshold and find analytically the number of intracavity emitters necessary for laser action to occur. We identify a universal route leading from thermal to coherent emission, through a collective anti-bunching regime that always precedes lasing. The coherent laser field always emerges  from a bifurcation with a well defined threshold  and has a frequency independent from the number of emitters.  \added[id=fp]{ Interferometric  measurements of the coherence time}~\cite{lebreton13a,ulrich07a} ($g^{(1)}(\tau)$) -- equivalent to detecting the existence of a well-defined frequency at the bifurcation -- \added[id=gd]{can} in principle be used to unambiguously identify lasing in all devices. This also holds for $\beta = 1$ (or close to this boundary), where the I/O characteristics  cannot carry useful information and measurements of $g^{(2)}(0)$ are poor indicators of lasing, leading to wrong conclusions on the existence or position of the laser threshold. The accurate characterization of the coherence properties of nanolasers holds promise for quantitative predictions to be used in a variety of applications, ranging from telecommunications and optical chips~\cite{Miller2017}, to the spectroscopic use of nanosources~\cite{Melentiev2017}, nanosensing~\cite{Stockman2015} and biophysical applications~\cite{Fikouras2018}.  These applications can greatly benefit from a detailed understanding and predictive power of designs where control in the number of quantum dots (now technologically feasible~\cite{Kaganskiy2019,Grosse2020}) coupled to effective coupling allow for tailored emission properties.  In addition, at the micro and nano scale this new  theory  can be used to investigate the interaction of emitters with nanostructures and arrays of particles~\cite{papoff11a,imura14a,mcarthur17b,mcarthur17c,mcarthur20a} and obtain predictions for new effects emerging from quantum interactions. 

The Strathclyde and UCA groups are grateful to the the CNRS and its Laboratoire International Associ\'e (LIA) ``Solace'' for support. The PhD of Mark Anthony Carroll is supported by the University of Strathclyde. GD and FP wish to thank the Mathematisches Forschungsinstitut Oberwolfach for support. GLL is grateful to L. Chusseau for discussions.


\end{document}


\preprint{APS/123-QED}

\title{Thermal, quantum anti-bunching and lasing thresholds from single  emitters to macroscopic devices}
\author{Mark Anthony Carroll}
\affiliation{Department of Physics, University
of Strathclyde,  107 Rottenrow,  Glasgow G4 0NG, UK.}
\author{Giampaolo D'Alessandro}
\affiliation{School of Mathematics, University of Southampton, Southampton SO17 1BJ, United Kingdom}
\author{Gian Luca Lippi}
\affiliation{Universit\'e C\^ote d'Azur, Institut de Physique de Nice, UMR 7710 CNRS, 1361 Route des Lucioles, 06560 Valbonne, France}
\author{Gian-Luca Oppo}\affiliation{Department of Physics, University
of Strathclyde,  107 Rottenrow,  Glasgow G4 0NG, UK.}
\author{Francesco Papoff}
\email{f.papoff@strath.ac.uk}
\affiliation{Department of Physics, University
of Strathclyde,  107 Rottenrow,  Glasgow G4 0NG, UK.}

\date{\today}
             
\maketitle

\section{The equations for non-identical emitters}

\added[id=fp]{The free energy and the light-matter interaction are given, in the Heisenberg picture, by the fully quantized Hamiltonian~\cite{chow14a}}

\begin{align}
    H = \hbar \sum_q \nu_q (b_q^\dagger b_q + \frac{1}{2}) + \sum_{q,n}[\varepsilon_{c,n} c_n^\dagger c_n + \varepsilon_{v,n}v_n^\dagger v_n - i\hbar(g_{nq}b_q c_n^\dagger v_n - g_{nq}^*b_q^\dagger v_n^\dagger c_n)],
\end{align}

\added[id=fp]{where $\nu_q$ is the frequency of the $q-th$ mode of the laser cavity, $b_q, b_q^{\dagger}$ are the destruction and creation operators for $q-th$ mode photons. $\varepsilon_{c,n}$ and $\varepsilon_{v,n}$
are the energies of the electors in the conduction and valence level in the $n-th$ quantum dot and $c_n, c_n^{\dagger}$ and $v_n, v_n^{\dagger}$ are creation and destruction operators, respectively, for conduction and valence electrons of the $n-th$ quantum dot. The equations for $\langle c_n^\dagger c_n\rangle$, $\delta \langle b_q c_n^\dagger v_n \rangle $, $ \delta \langle b^\dagger_q b_q \rangle$, $\langle b_q\rangle$,  and
$ \langle v_n^\dagger c_n \rangle$ are:}

\begin{eqnarray}
&&    d_t  \langle c_l^\dagger c_l\rangle = - \gamma_{nr} \langle c_l^\dagger c_l\rangle + r(1 - \langle c_l^\dagger c_l\rangle) - 2\sum_q \mathfrak{Re}g_{ql}(\delta\langle b_q c_l^\dagger v_l\rangle + \langle b\rangle\langle c_l^\dagger v_l\rangle),
\\
&&    d_t  \delta\langle b_q c_l^\dagger v_l\rangle = -[\gamma_q + \gamma +i(\nu_q - \Delta\varepsilon_l)]\delta\langle b c_l^\dagger v_l\rangle + g_{ql}^*[\langle c_l^\dagger c_l\rangle + \delta\langle b^\dagger_q b_q\rangle(2\langle c_l^\dagger c_l\rangle - 1) - \langle c_l^\dagger v\rangle\langle v_l^\dagger c_l\rangle ],
\\
&&    d_t  \delta\langle b^\dagger_q b_q \rangle = -2\gamma_q \delta\langle b^\dagger_q b_q \rangle + 2\sum_l\mathfrak{Re}g_{ql}\delta\langle b_q c_l^\dagger v_l\rangle,
\\
&&    d_t  \langle b_q \rangle = -(\gamma_q + i\nu_q)\langle b_q \rangle + g_{ql}^*N\langle v_l^\dagger c_l\rangle,
\\
&&    d_t  \langle v_l^\dagger c_l\rangle = -(\gamma + i\Delta\varepsilon_l)\langle v_l^\dagger c_l\rangle + 
\sum_q g_{ql}[ \langle b_q \rangle(2\langle c_l^\dagger c_l\rangle - 1)],
\end{eqnarray}

\added[id=fp]{where the dissipative terms are obtained with a Lindblad formalism\cite{florian13a, leymann14a}. Neglecting intensities and field amplitudes of the non resonant modes $q \ne 0$ and eliminating adiabatically the correlations $\delta\langle b_q c_l^\dagger v_l\rangle$, we find that non resonant modes can be accounted for by  the additional decay rate for the population $\gamma_{nr}$. Assuming that only one mode is resonant with the cavity and the $N$ emitters coupled to the laser mode are identical~\cite{kreinberg17a,moody18a},  leads to the simpler equations given to Eqs(1-5) in the main paper, which give the dynamic of the system after a temporal transient in which the emitters' variables are  equalized.}

We present the analytic results of the non-lasing and lasing solutions. The following notation is used throughout: $\Gamma_c = \gamma + \gamma_c$ , \; $\Delta_\Gamma = \gamma_c (\Gamma_c^2+\Delta\nu^2)$, \; $\Gamma_n = \gamma_{nl}+\gamma_{nr}$, $\Gamma_t=3\gamma_c + \gamma$ and $\Delta\Omega = \Delta\nu(1 - \frac{\gamma_c}{\Gamma_c})$. Parameter definitions are: light-matter coupling, $g$; cavity decay rate, $\gamma_c$; polarisation dephasing, $\gamma$; non-radiative decay, $\gamma_{nr}$; non-lasing decay, $\gamma_{nl}$; detuning, $\Delta\nu$; and the number of emitters, $N$. 

All of the figures in the main paper and in this supplementary material have been obtained analytically with the non-lasing and lasing solutions. These have been verified numerically using MATLAB and Maple. 

\section{Numerical tests}

\begin{figure}
  \centering
  \includegraphics[width=0.5\columnwidth]{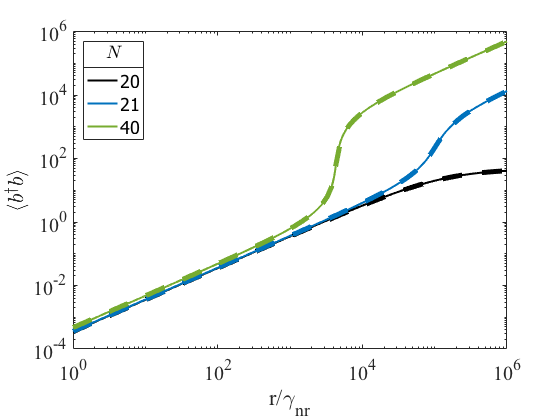}
  \caption{Comparison of the analytic (solid lines) and numerical (dashed lines) results of the intensity versus pump rate $r$ for different numbers of emitters. Parameters: $\beta=7\times10^{-4}$, $\gamma_{nl}=1.4\times10^{12}s^{-1}$, $\gamma_{l}=9.68\times10^{8}s^{-1}$,  $\gamma=10^{13}s^{-1}$, $\gamma_c=10^{10}s^{-1}$, $\gamma_{nr}=10^9s^{-1}$ and $g=7\times10^{10}s^{-1}$. }
  \label{fig:AnalyticVsNumerical}
\end{figure}

\added[id=mc]{A comparison between the analytic solutions and the numerical steady state values in Fig.~\ref{fig:AnalyticVsNumerical} shows a perfect agreement indicating that the numerical simulations are on firm ground. }

\begin{figure}[hbt!]
  \subfloat[\label{subfig:a}]{%
    \includegraphics[width=0.5\columnwidth]{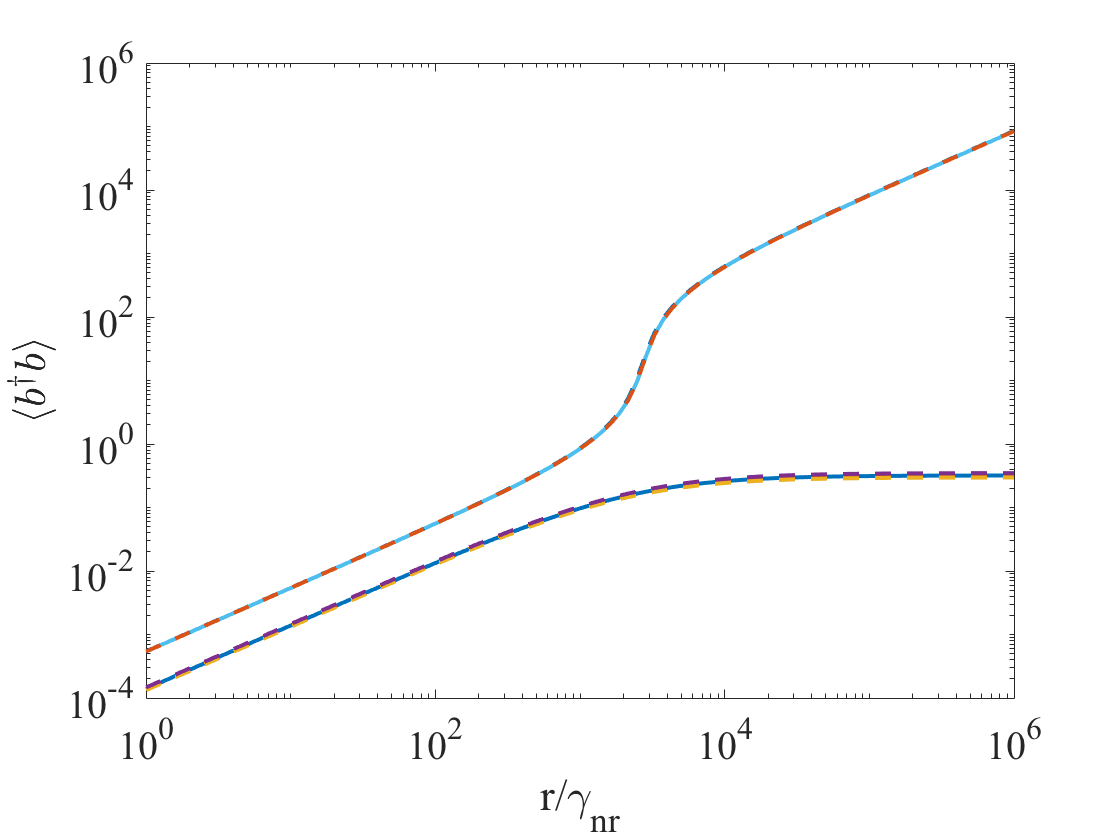}%
  }
  \caption{Intensity as a function of the pump rate for five quantum dots where in the lower branch $g=7\times10^{10}s^{-1}$ ($\beta=7\times10^{-4}$) and in the upper branch $g=25\times10^{10}s^{-1}$ ($\beta=9\times10^{-3}$) .  The dashed lines are obtained from simulations where $g$ and $\Delta\nu$ are varied randomly up to 10\% of the values used to obtain the solid lines. Parameters: $\gamma=10^{13}s^{-1}$, $\gamma_c=10^{10}s^{-1}$, $\gamma_{nr}=10^9s^{-1}$, $\gamma_{nl}=1.4\times10^{12}s^{-1}$ and $\Delta\nu=10^{12}s^{-1}$.  }
  \label{fig:Intensity}
\end{figure}

\added[id=mc]{Following on from this in Fig.~\ref{fig:Intensity} we give the steady state solutions of the photon number versus the pump rate $r$ for two cases: the first assumes all emitters to be identical and as such only requires one equation for each variable for $N$ emitters, Eqs. (1-5) in the main paper (solid lines); and in the second case we modify the model to include $N$ equations for $\langle c^\dagger c\rangle$, $\langle v^\dagger c\rangle$ and $\delta\langle b c^\dagger v\rangle$ (dashed lines). In the latter case, we include values for the detuning $\Delta\nu$ and coupling coefficient $g$ that are varied by up to 10\% of the values used to obtain the solid curves (the random numbers used were drawn from a uniform distribution). In both cases, which are initiated with random initial conditions, after a temporal transient the variables steady state values converge establishing a collective state.   } 

\section{Comparison with rate and master equation models}

\begin{figure}[hbt!]
  \subfloat[\label{subfig:3a}]{%
    \includegraphics[width=0.5\columnwidth]{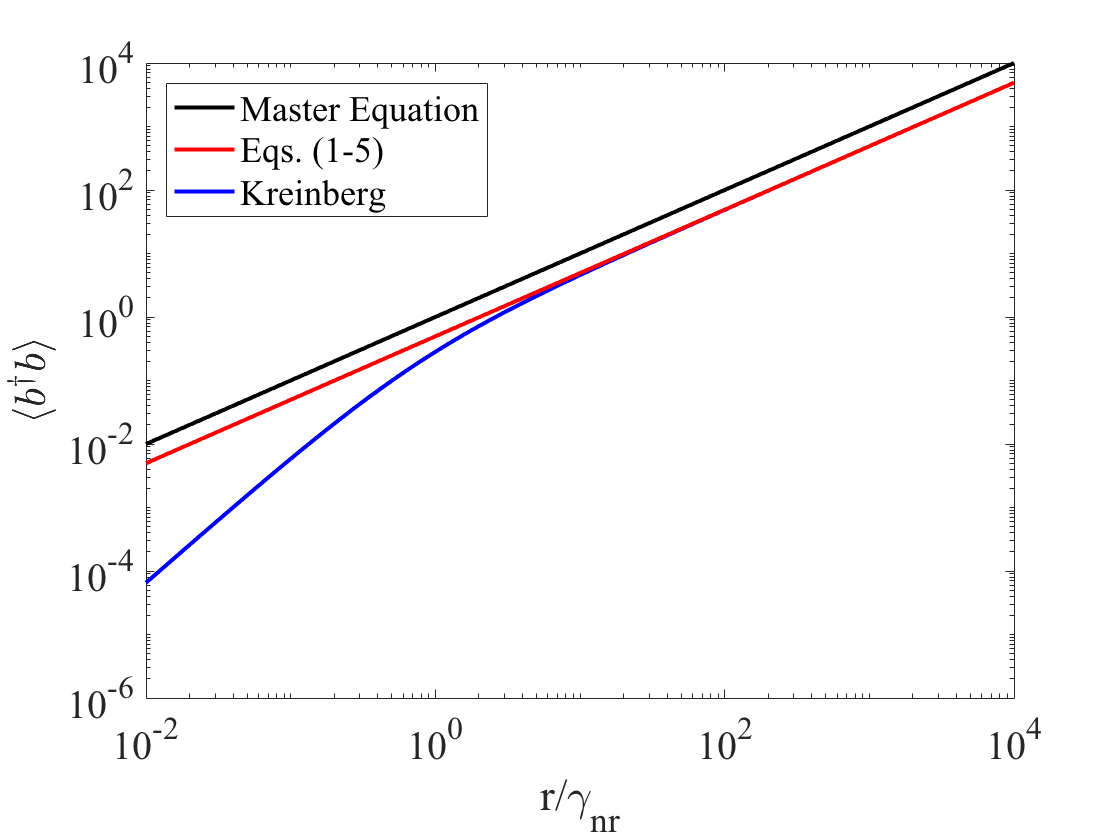}%
  }
  \subfloat[\label{subfig:3b}]{%
    \includegraphics[width=0.5\columnwidth]{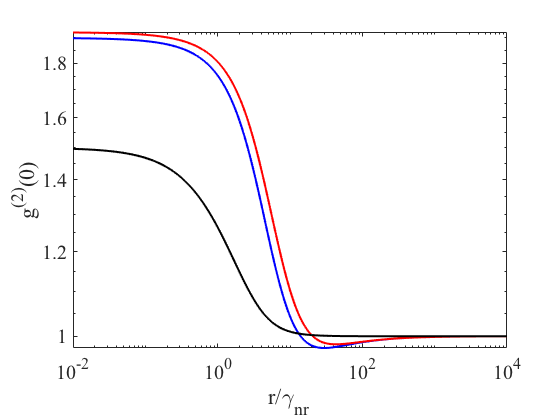}%
  }
  \caption{Comparison of the intensity between a fully quantized master equation and Eqs. (1-5) in the main paper where $\beta=1$. Parameters:  $g=7\times10^{10}s^{-1}$, $\gamma=10^{13}s^{-1}$, $\gamma_c=10^{10}s^{-1}$, $\gamma_{nr}=10^9s^{-1}$ and $N=40$.}
  \label{fig:MasterEquationComparison}
\end{figure}

\added[id=mc]{\added[id=gd]{We make here} a direct comparison between Eqs. (1-5) in the main paper with a fully quantized master equation \cite{rice1994photon} and a cQED model \cite{kreinberg17a} for the case of $\beta=1$ . Under the condition of $\beta=1$ the results of the master equation are calculated via Eq. 8a and Eq. 36 from \cite{rice1994photon}. The main difference between Eqs. (1-5) and the master equation given in \cite{rice1994photon} is that the master equation derives from the two rate equation variables, intensity and excited carrier number, and does not include the medium polarisation. Without the polarisation there are only correlations between the carrier number and intensity, whereas we have correlations between the field and polarisation of the medium which leads to the anti-bunching. Fig.~\ref{fig:MasterEquationComparison} shows a comparison of the intensity obtained through Eqs. (1-5) of the main paper with a fully quantized master equation. The divergence between the red and the blue curves in Fig.~\ref{subfig:3a} for low pump rate can be accounted for by noting that in \cite{kreinberg17a} the term responsible for spontaneous emission is nonlinear whereas in Eq. (2) of the main paper it is linear resulting from the assumption of two-level emitters.   }

\begin{figure}[hbt!]
  \centering
  \includegraphics[width=0.5\columnwidth]{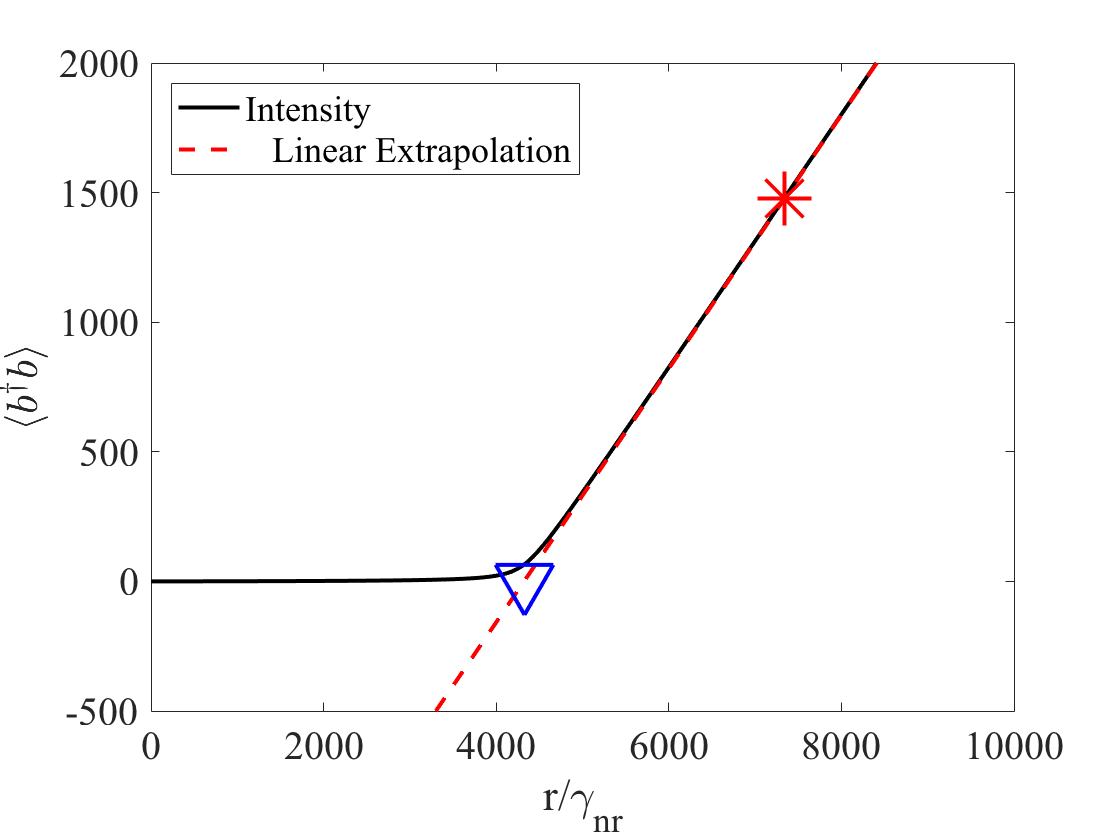}
  \caption{Linear plot of intensity versus pump rate $r$ (black) and the extrapolated fit (red) where $\beta=7\times10^{-4}$. The red star marks the position of the lasing bifurcation and the blue triangle shows where the linear extrapolation of the intensity crosses zero. Parameters:  $g=7\times10^{10}s^{-1}$, $\gamma=10^{13}s^{-1}$, $\gamma_c=10^{10}s^{-1}$, $\gamma_{nr}=10^9s^{-1}$, $\gamma_{nl}=1.4\times10^{12}s^{-1}$ and $N=40$.}
  \label{fig:Extrapolation}
\end{figure}

\added[id = mc]{In Fig.~\ref{fig:Extrapolation} we show the intensity versus pump above and below threshold. Using the usual procedure of linear extrapolation of the intensity down to zero and comparing the value of the laser threshold with that obtained from Eqs. (1-5) in the main paper. The red star marks the position of the lasing threshold calculated from Eqs. (1-5) from the main paper and the blue triangle marks where the linear extrapolation of the intensity crosses zero. There are two points to note: the value of the laser threshold predicted from linear extrapolation occurs at the point of initial rapid growth of photons; and, the position of the red star occurs after this rapid growth in photon number.}

\begin{figure}[hbt!]
  \centering
  \includegraphics[width=0.5\columnwidth]{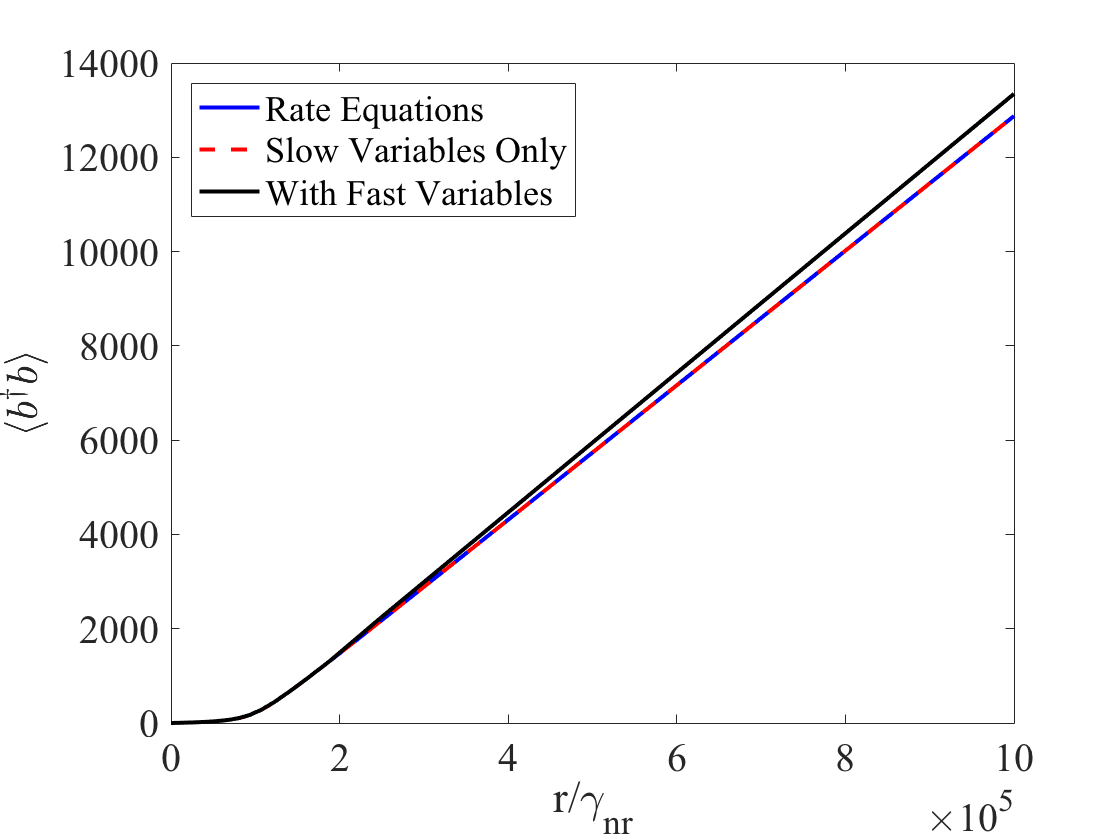}%
  \caption{Comparison of the intensity versus pump for three cases: rate equations; the model with only slow variables; and the model with slow and fast variables. Parameters:  $g=7\times10^{10}s^{-1}$, $\gamma=10^{13}s^{-1}$, $\gamma_c=10^{10}s^{-1}$, $\gamma_{nr}=10^9s^{-1}$, $\gamma_{nl}=1.4\times10^{12}s^{-1}$ and $N=21$. The equations corresponding to ignoring the fast variables are Eqs. (1-3) in the main paper (with the fast variables Eqs. (4-5) from the main paper are included).}
  \label{fig:3models}
\end{figure}

\added[id=mc]{Finally,  in Fig.~\ref{fig:3models} we compare the intensity versus pump highlighting the differences between Eqs. (1-5) in the main paper with and without the fast variables and rate equations derived from Eqs. (1-5). In order to derive rate equations we consider only the slow variables and insert the adiabatic solution of the photon assisted polarisation into the equations for the carrier population and intensity.} \added[id=gd]{The rate equations and the model with only the slow variables produce almost identical values of the intensity.  The inclusion of the fast variables changes slightly the intensity, with the change being more significant for larger values of the decay of the medium polarisation, $\gamma$, with respect to the cavity decay rate, $\gamma_{c}$.  The key role that the fast variables play, is that without them it is not possible neither to define a laser threshold nor to determine the well defined lasing frequency that is established after this bifurcation.}

\section{Analytic non-lasing solutions}
The photon number is given solely in terms of the system parameters:
\begin{eqnarray}
    \label{Eq1}
 \delta \langle b^\dagger b\rangle =  \frac{1}{8|g|^2\gamma_c \Gamma_c}
 \{-(\Gamma_n + r)\Delta_\Gamma-|g|^2 \Gamma_c[2\gamma_c +
 N(\Gamma_n-r) ] + F^{1/2} \}  
\end{eqnarray}
where $F$ is 
\begin{eqnarray}
    \label{Eq2}
   F = (N\Gamma_c\vert g\vert^2+\Delta_\Gamma)^2\Gamma_n^2-2(N\Gamma_c|g|^2+\Delta_\Gamma)((N\Gamma_c|g|^2-\Delta_\Gamma)r-2|g|^2\gamma_c\Gamma_c)\Gamma_n \nonumber \\ +(N\Gamma_c|g|^2-\Delta_\Gamma)^2r^2 
     +12\gamma_c\Gamma_cg^2(N\Gamma_c|g|^2+\frac{\Delta_\Gamma}{3})\alpha + 4|g|^4\gamma_c^2\Gamma_c^2.
\end{eqnarray}
The carrier density can then be given as a function of the photon number:
\begin{equation}
\label{Eq3}
{\langle c^\dagger c\rangle}={\frac {1}{\Gamma_n+r} \left( r-2\,{\frac {\gamma_{{c
}}\delta { \langle b^\dagger b\rangle}}{N}} \right) }.
\end{equation}
Of the three fundamental variables within the non-lasing solutions the photon polarisation is the only complex variable and we give it in terms of the carrier density $\langle c^\dagger c\rangle$ and photon number $\delta\langle b^\dagger b\rangle$
\begin{equation}
\label{Eq4}
\delta {\langle b c^\dagger v\rangle}=\frac{g^*(\Gamma_c-i\Delta\nu)}{\Gamma_c^2+\Delta\nu^2}\bigg[\langle c^\dagger c\rangle+\delta\langle b^\dagger b \rangle(2\langle c^\dagger c\rangle-1)\bigg].
\end{equation}
The equation for $g^{(2)}(0)$ is given by
\begin{align}
\label{Eq5}
   g^{(2)}(0)= 2+\frac{\delta\langle b^\dagger b^\dagger bb\rangle}{\delta\langle b^\dagger b\rangle^2}
\end{align}
where $A=\delta\langle b^\dagger b^\dagger bb\rangle $ and $B=\delta\langle b^\dagger b\rangle^2$ from the main paper. As seen from Eq.~\ref{Eq5}, in order to calculate $g^{(2)}(0)$ one must solve higher order correlation functions up to the four-particle level of the type $\delta\langle b^\dagger b^\dagger bb\rangle$. These higher order correlations couple to the photon assisted polarisation through the three-particle carrier-photon correlation $\delta\langle b^\dagger b c^\dagger c\rangle$. As an approximation we neglect correlations greater than two-particles from the dynamics of $\langle c^\dagger c\rangle$, $\delta\langle b^\dagger b\rangle$ and $\delta\langle b c^\dagger c\rangle$. This turns out to be a very good approximation as can be seen from the purple triangle markers in Fig. 1 of the main paper. With the aforementioned approximation we find analytically the solution to $\delta\langle b^\dagger b^\dagger bb\rangle$ as a function of the carrier density and photon number  
\begin{align}
\label{Eq6}
    \delta\langle b^\dagger b^\dagger bb\rangle = \frac{C}{D}
\end{align}
where
\begin{align}
\label{Eq7}
    D =\{(2 \gamma_c + \gamma_{nr})[Ng^2 \Gamma_t (2 \langle c^\dagger c\rangle -1) -\gamma_c(\Delta\nu^2+\Gamma_t^2)] -4g^2\Gamma_t \gamma_c (2\delta\langle b^\dagger b\rangle+1) \} (\Delta\nu^2 + \Gamma_c^2)^2
\end{align} 
and 
\begin{align}
\label{Eq8}
    C =4|g|^2\Gamma_t\bigg\{4|g|^2 N \Gamma_c^2( 2\gamma_c   + \gamma_{nr})( \delta\langle b^\dagger b\rangle^2\langle c^\dagger c\rangle^2 +  \delta\langle b^\dagger b\rangle \langle c^\dagger c\rangle^2)  +[2\gamma_c(  \Gamma_c^2   +  \Delta\nu^2 )^2 \nonumber \\ - 4|g|^2N\Gamma_c^2(2\gamma_c  + \gamma_{nr})] \delta\langle b^\dagger b\rangle^2 \langle c^\dagger c\rangle  + |g|^2 N \Gamma_c^2 (2\gamma_c  + \gamma_{nr}) \langle c^\dagger c\rangle^2    + 2\gamma_c(\Delta\nu^2 \nonumber \\+ \Gamma_c^2)^2 \delta\langle b^\dagger b\rangle^3  +[ \gamma_c ( \Gamma_c^2+ \Delta\nu^2)^2 +|g|^2N\Gamma_c^2 ( 2\gamma_c   + \gamma_{nr})] \delta\langle b^\dagger b\rangle^2 \nonumber \\ +[-2|g|^2 N \Gamma_c^2(2\gamma_c  + \gamma_{nr})  +\gamma_c (\Delta\nu^2 + \Gamma_c^2)^2] \delta\langle b^\dagger b\rangle \langle c^\dagger c\rangle  \bigg\} 
\end{align}

\section{analytic laser solutions}
The {\it Modulus of} the Coherent field (solution exists only when $| \langle b \rangle | \ge 0$) is given by
\begin{align}
\label{Eq9}
|\langle b \rangle | = 
\frac{1}{ \gamma |g|\Delta_\Gamma} \bigg[\frac{( N|g|^2\Gamma_c^2 -\gamma\Delta_\Gamma)\Delta_\Gamma r -( N|g|^2\Gamma_c^2 +  \gamma \Delta_\Gamma ) ( \Delta_\Gamma \Gamma_n +2|g|^2\Gamma_c^2)}{4(N-1)}(\gamma^2 + \Delta\Omega^2)N \bigg]^{\frac{1}{2}}
\end{align}
where the expectation value of the coherent field is defined as $\langle b\rangle = |\langle b\rangle|e^{-i( \Omega t +  \phi)}$; the phase $\phi$ is undetermined and there are in principle an infinite set of solutions. Note that for $\Omega\neq0$ the bifurcation is a Hopf bifurcation; however, $e^{-i\Omega t}$ can be factored out and the bifurcation reduced to a pitchfork bifurcation.  The steady state solution of the carrier density is
\begin{equation}
\label{Eq10}
    \langle c^\dagger c\rangle = \frac{1}{2}\bigg(1 + \frac{\gamma\Delta_{\Gamma}}{N|g|^2\Gamma_c^2}\bigg)
\end{equation}
and the standard polarisation of the medium is defined as
\begin{equation}
\label{Eq11}
    \langle v^\dagger c\rangle = \frac{g}{\gamma + i{\Delta\Omega}}\langle b\rangle(2\langle c^\dagger c\rangle-1).
\end{equation}
The solution of the intensity correlation is 
\begin{equation}
\label{Eq12}
 \delta\langle b^\dagger b\rangle = \frac{N}{2\gamma_c}\Bigg[r - (\Gamma_n + \alpha)\langle c^\dagger c\rangle - \frac{
 2\vert g \vert^2\gamma}{\gamma^2+\Delta\Omega^2}\vert \langle b \rangle\vert^2(2\langle c^\dagger c\rangle-1)\bigg],
\end{equation}
and the solution to the photon assisted polarisation is
\begin{align}
\label{Eq13}
    \delta\langle b c^\dagger v\rangle = \frac{g^*}{\Gamma_c + i\Delta\nu}\bigg[\langle c^\dagger c\rangle + \delta\langle b^\dagger b\rangle(2\langle c^\dagger c\rangle-1)-\vert\langle v^\dagger c\rangle\vert^2\bigg].
\end{align}

\section{The value of the pump at the laser threshold }

Solving Eq.~\ref{Eq9} for the pump $r$ at threshold, $|\langle b \rangle|=0$, we get the following equation 
\begin{align}
\label{Eq14}
    r \ge \frac{(N|g|^2\Gamma_c^2 + \gamma\Delta_{\Gamma}){(2|g|^2\Gamma_c^2 + \Delta_{\Gamma}\Gamma_n})}{\Delta_{\Gamma}(N|g|^2\Gamma_c^2 - \gamma\Delta_{\Gamma})}
\end{align}
imposing the physical conditions that $r$ is positive and the denominator of Eq.~\ref{Eq14} cannot be zero. When $r$ is equal to the term on the right hand side in Eq.~\ref{Eq14} marks the lasing threshold. We find the condition that satisfies the physical constraints of $r$ to be the inequality 
\begin{align}
\label{Eq15}
    N > \frac{\gamma\Delta_{\Gamma}}{|g|^2\Gamma_c^2}
\end{align}
which is the same condition given in the main paper that states for a lasing threshold to exist there must be a critical number of emitters. If the condition in Eq.~\ref{Eq15} is fulfilled then there is the solution for $|\langle b \rangle|$ and it grows in the shape of a pitchfork bifurcation.  
\section{Phase diagrams with detuning}
\begin{figure}[ht!]
\subfloat[\label{subfig:6a}]{%
  \includegraphics[width=0.3\columnwidth]{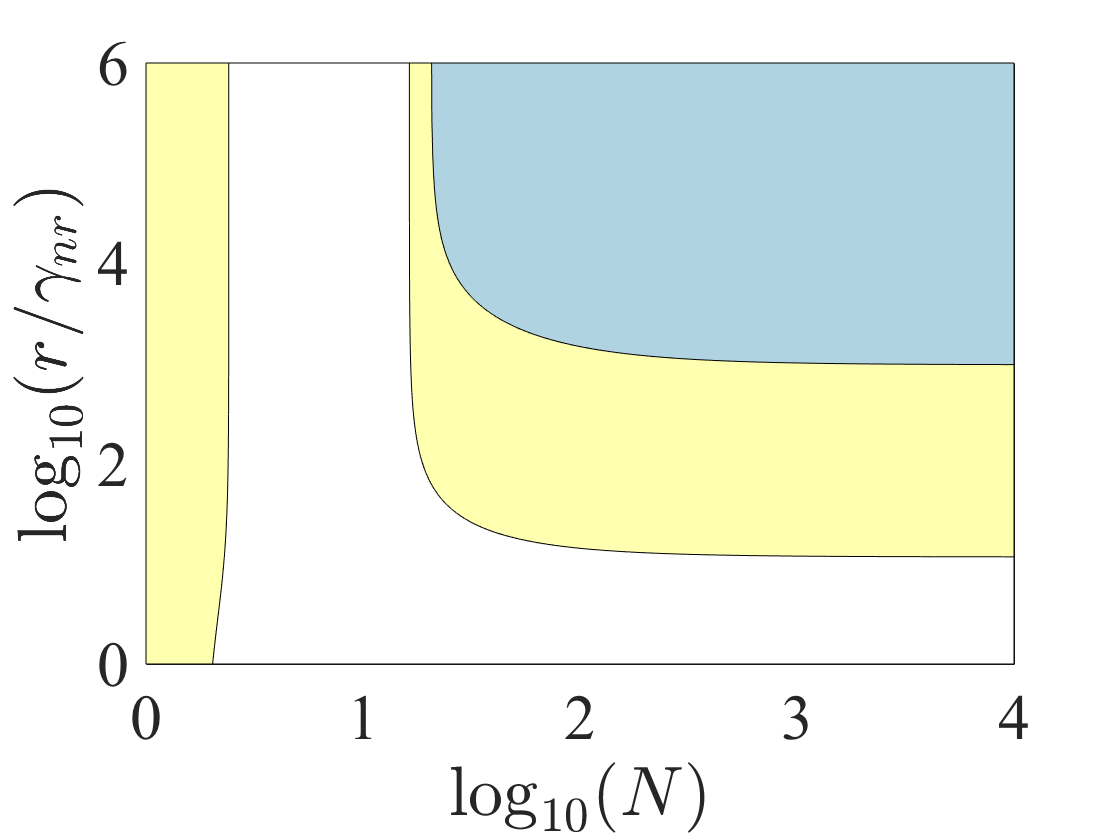}%
}
\subfloat[\label{subfig:6b}]{%
  \includegraphics[width=0.3\columnwidth]{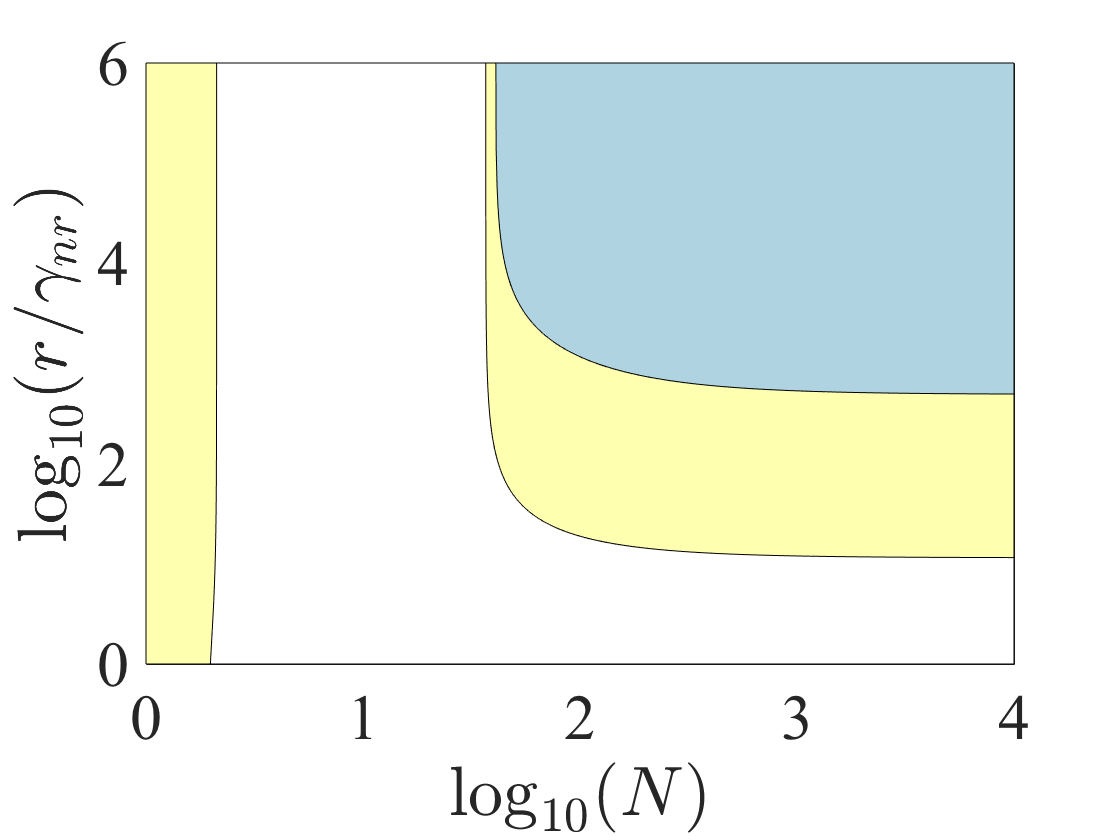}%
}
\subfloat[\label{subfig:6c}]{%
  \includegraphics[width=0.3\columnwidth]{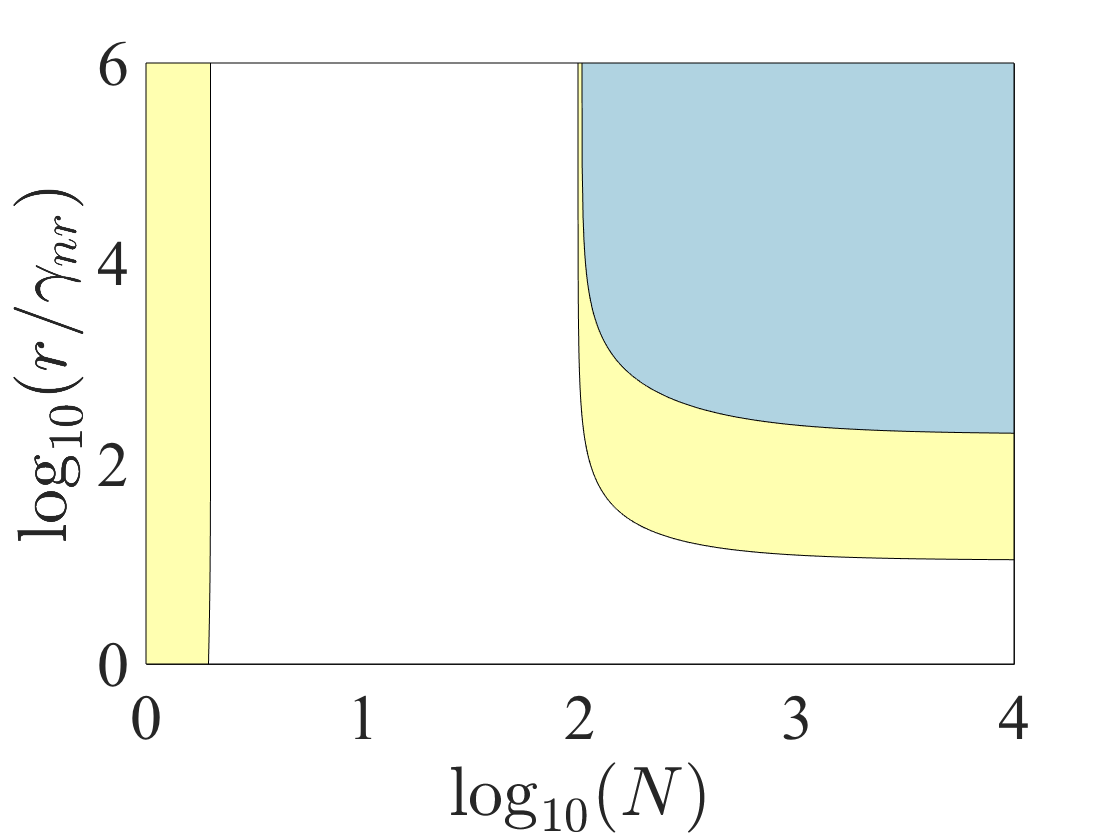}%
}\hfill
\subfloat[\label{subfig:6d}]{%
  \includegraphics[width=0.3\columnwidth]{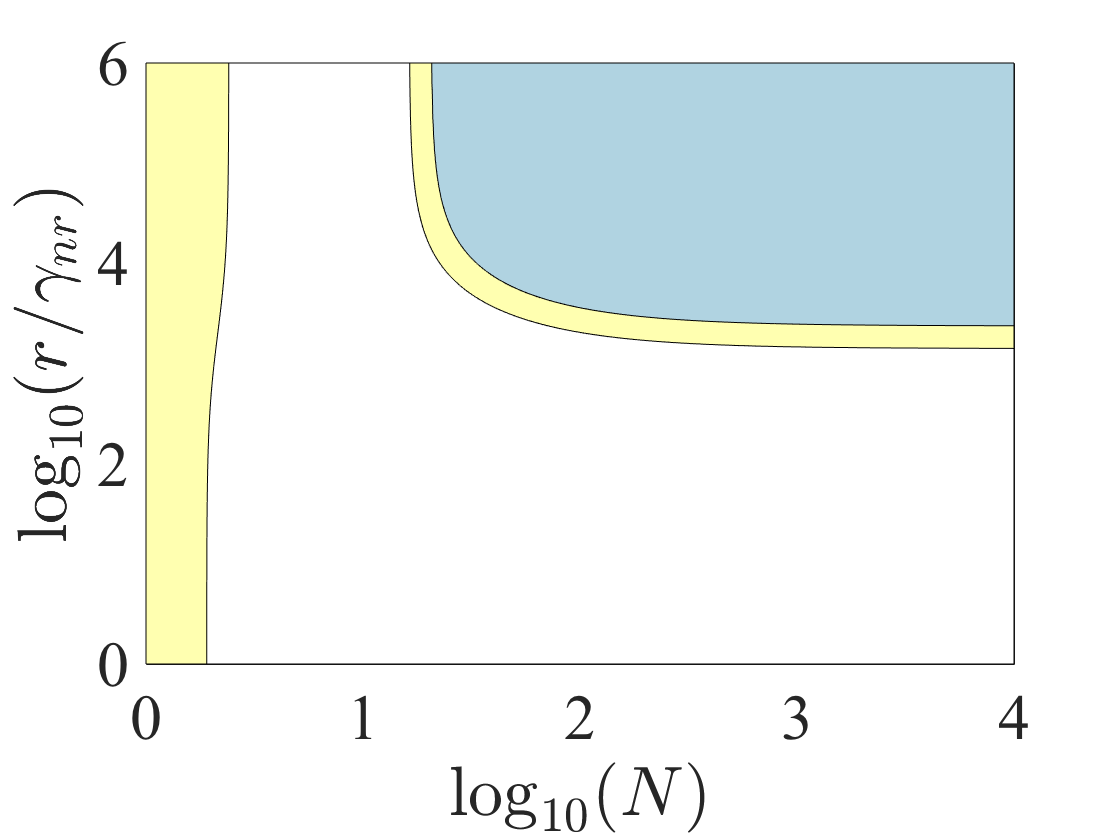}%
}
\subfloat[\label{subfig:6d}]{%
  \includegraphics[width=0.3\columnwidth]{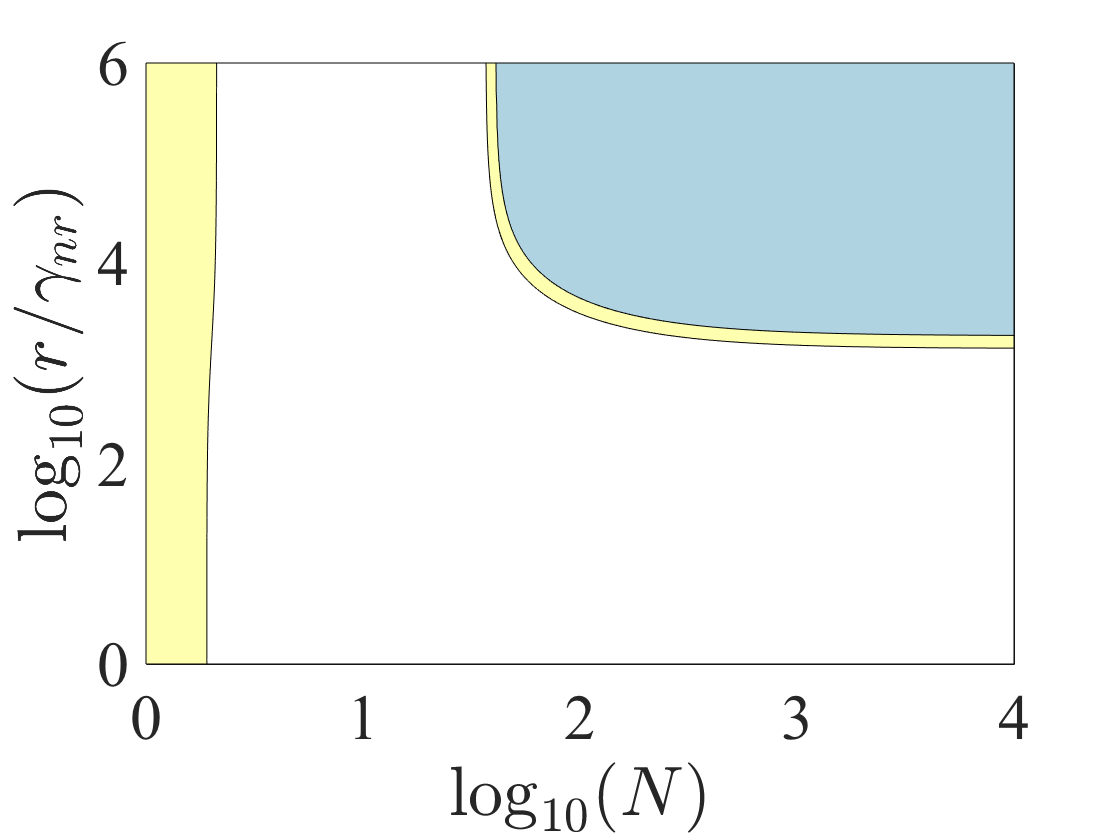}%
}
\subfloat[\label{subfig:6d}]{%
  \includegraphics[width=0.3\columnwidth]{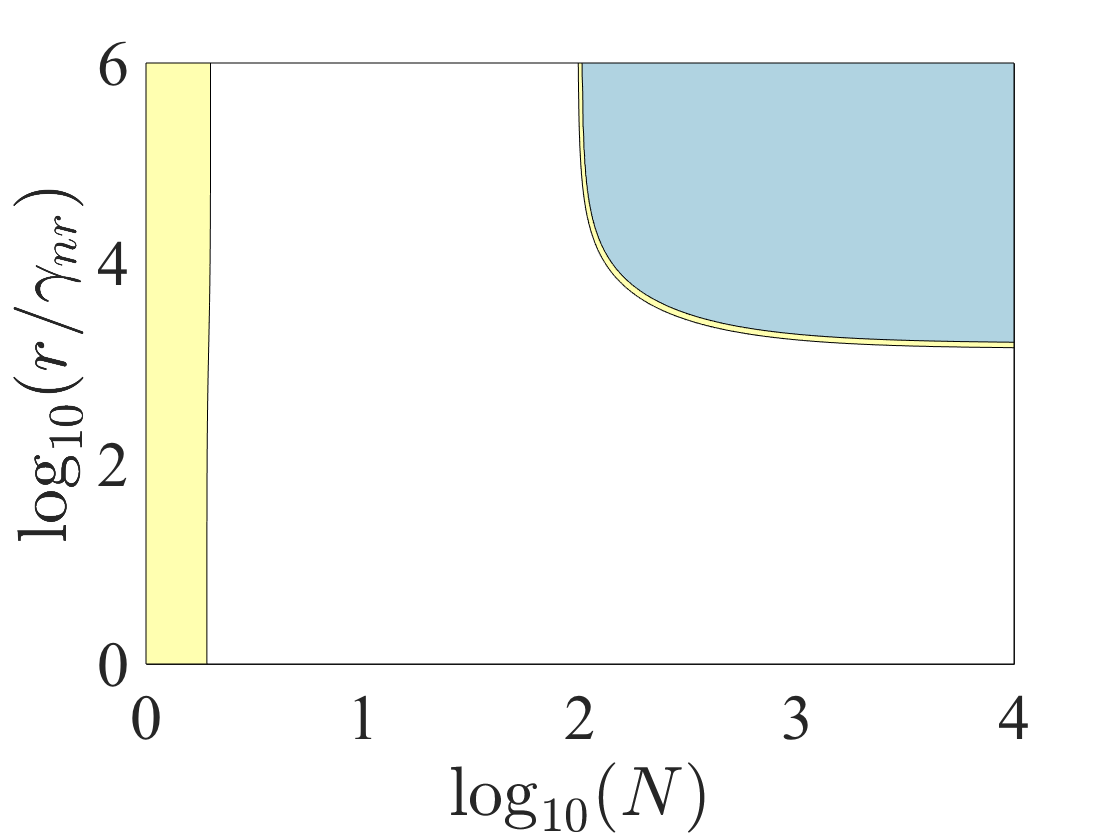}%
}
\caption{Phase diagrams for devices with changing detuning and spontaneous emission factor $\beta=1$ (a)-(c), and  $7\times10^{-4}$ (d)-(f). The detunings are in (a) and (d),  $\Delta\nu=10^{12}s^{-1}$; (b) and (e),  $\Delta\nu=10^{13}s^{-1}$; (c) and (f),  $\Delta\nu=2\times10^{13}s^{-1}$. Common to both values of $\beta$ is the increase of $N_c$ as the detuning increases, and in both cases we find that the extent of the anti-bunching regime as a function of the pump is reduced. Parameters: $\gamma=10^{13}s^{-1}$, $\gamma_c=10^{10}s^{-1}$, $\gamma_{nr}=10^9s^{-1}$, and $g=7\times10^{10}s^{-1}$. The white region corresponds to the thermal regime; yellow to the anti-bunching regime; and blue to the lasing regime.}\label{fig:PhaseDiagrams}
\end{figure}

In Fig.~\ref{fig:PhaseDiagrams} the affect of detuning is shown for devices with different values of $\beta$. In both cases we find that increasing the detuning shifts the value of $N_c$ towards larger values of $N$. This is confirmed by inspecting the numerator in Eq.~\ref{Eq15} where increasing the detuning results in a larger value of $N$ to satisfy the physical conditions put in place for $r$.

\section{Linear stability analysis}

\added[id=fp]{With the vector notation 
$\underline{i}=(\langle c^\dagger c\rangle,\delta\langle b^\dagger b\rangle,\delta\langle b c^\dagger v\rangle, \delta\langle b^\dagger v^\dagger c\rangle)$, $\underline{c} = (\langle b\rangle,\langle v^\dagger c\rangle,\langle b^\dagger \rangle,\langle c^\dagger v\rangle)$, the model equations become
$d_t \underline{i} = \underline{F}(\underline{i},\underline{c}), \;\; d_t \underline{c} = \underline{G}(\underline{i},\underline{c})$, where $\underline{F}(\underline{i},\underline{c})$ and $\underline{G}(\underline{i},\underline{c})$ are non linear vector functions of $\underline{i}$ and $\underline{c}$ whose components are the right-hand side of Eqs.(1-5) and their complex conjugate. For instance, 
$G_{\langle b\rangle}(\tilde{\underline{i}},\tilde{\underline{c}})$ and
$G_{\langle v^\dagger c\rangle}(\tilde{\underline{i}},\tilde{\underline{c}})$ are the left-hand sides of the equations for $d_t \langle b\rangle$ and $d_t \langle v^\dagger c\rangle$, respectively,  evaluated at the solution $\tilde{\underline{i}},\tilde{\underline{c}}$. The dynamics in 
the linear regime of small perturbations $\eta_i, \eta_c$ of an arbitrary solution $\tilde{\underline{i}},\tilde{\underline{c}}$ of Eqs.(1-5) is given by}

\begin{equation}
d_t \left[ \begin{array}{c}  \underline{i} \\ \underline{c} \end{array} \right] = 
\left[ \begin{array}{cc} \nabla_i \otimes \underline{F}(\tilde{\underline{i}},\tilde{\underline{c}}) & \nabla_c \otimes \underline{F}(\tilde{\underline{i}},\tilde{\underline{c}}) \\ 
\nabla_i \otimes \underline{G}(\tilde{\underline{i}},\tilde{\underline{c}}) & \nabla_c \otimes \underline{G}(\tilde{\underline{i}},\tilde{\underline{c}}) \end{array} \right] \left[ \begin{array}{c}  \underline{i} \\ \underline{c} \end{array} \right] \label{gen_ls}
\end{equation}

\added[id=fp]{where each block in matrix on the right-hand side of Eq.~\eqref{gen_ls} is of dimension $4 \times 4$. For any solution with $\tilde{\underline{c}}=0$ one has
$\nabla_i \otimes \underline{G}(\tilde{\underline{i}},0)=0$, so the matrix in Eq.~\eqref{gen_ls} is in block-triangular firm and it's eigenvalues are the eigenvalues of the block along the main diagonal. Furthermore, the perturbations of $\langle b\rangle,\langle v^\dagger c\rangle$ and their complex conjugate are decoupled so that}

\begin{equation}
\nabla_c \otimes \underline{G}(\tilde{\underline{i}},0) =
\left[ \begin{array}{cc}
J & 0 \\ 
0 & J^\dagger \end{array} \right], 
\end{equation}

\added[id=fp]{where $J$ is the $2 \times 2$ matrix} 

\begin{equation}
J = \left[ \begin{array}{cc}
\partial_{\langle b\rangle}G_{\langle b\rangle}(\tilde{\underline{i}},0) & \partial_{\langle v^\dagger c\rangle} G_{\langle b\rangle}(\tilde{\underline{i}},0)\\ 
\partial_{\langle b\rangle} G_{\langle v^\dagger c\rangle}(\tilde{\underline{i}},0) & \partial_{\langle v^\dagger c\rangle} G_{\langle v^\dagger c\rangle}(\tilde{\underline{i}},0) \end{array} \right].
\end{equation}

\added[id=mc]{All data underpinning this publication are openly available from the University of Strathclyde KnowledgeBase at https://doi.org/10.15129/eb90c359-cf7c-4ce5-a40f-770092f4ef63.}
